\documentclass[sigconf, authorversion]{acmart}

\usepackage{enumitem}
\setlist[itemize]{leftmargin=*}
\setlist[enumerate]{leftmargin=*,label=\arabic*.}

\usepackage[shortcuts]{extdash}

\usepackage{soul}

\usepackage{xspace}

\usepackage{xparse}

\newcommand{\code}[1]{{\small{\texttt{#1}}}}

\NewDocumentCommand\p{ m g }{%
    \IfNoValueTF{#2}{%
         {\small\mbox{$p<#1$}}%
    }{%
         {\small\mbox{$p#2#1$}}%
    }%
}

\NewDocumentCommand\anova{ m m m m g }{%
    \IfNoValueTF{#5}{%
         {\small$F_{#1,#2}=#3$, $p<#4$}%
    }{%
         {\small$F_{#1,#2}=#3$, $p<#4$, $\eta_G^2=#5$}%
    }%
}

\newenvironment{commentwrapper}[1]{\color{#1}}{\color{black}}

\definecolor{GREEN}{rgb}{0.0,0.7,0.0}
\definecolor{BLUE}{rgb}{0.0,0.2,0.7}
\definecolor{GOLD}{rgb}{0.6,0.6,0.0}
\definecolor{CYAN}{rgb}{0.0,0.6,0.6}
\definecolor{PURPLE}{rgb}{0.6,0.0,0.6}

\definecolor{RED}{rgb}{0.7,0.0,0.0}
\definecolor{GRAY}{gray}{0.5}

\definecolor{LIGHTGRAY}{HTML}{DDDDDD}

\usepackage{amsmath}
\usepackage{multirow}
\usepackage{subcaption}
\usepackage{wrapfig}
\usepackage{epigraph}
\usepackage{graphics}
\usepackage{acmart-taps}

\usepackage{csquotes}

\colorlet{lightgray}{gray!20!}

\usepackage{cuted}
\usepackage{capt-of}

\usepackage[rightmargin=0cm,vskip=0pt,noorphans,font={itshape},indentfirst=false]{quoting}

\AtBeginDocument{%
  \providecommand\BibTeX{{%
    \normalfont B\kern-0.5em{\scshape i\kern-0.25em b}\kern-0.8em\TeX}}}

\copyrightyear{2022}
\acmYear{2022}
\setcopyright{acmlicensed}
\acmConference[UIST '22]{The 35th Annual ACM Symposium on User Interface Software and Technology}{October 29-November 2, 2022}{Bend, OR, USA}
\acmBooktitle{The 35th Annual ACM Symposium on User Interface Software and Technology (UIST '22), October 29-November 2, 2022, Bend, OR, USA}
\acmPrice{15.00}
\acmDOI{10.1145/3526113.3545617}
\acmISBN{978-1-4503-9320-1/22/10}

\begin{document}

\title[Code-Driven Storytelling]{CodeToon: Story Ideation, Auto Comic Generation, and Structure Mapping for Code-Driven Storytelling}

\author{Sangho Suh, Jian Zhao, Edith Law}
\affiliation{
  \country{University of Waterloo, Canada}
  }
\email{{sangho.suh, jianzhao, edith.law}@uwaterloo.ca}




\renewcommand{\shortauthors}{Sangho Suh, Jian Zhao, Edith Law}

\begin{abstract}
Recent work demonstrated how we can design and use coding strips, a form of comic strips with corresponding code, to enhance teaching and learning in programming. However, creating coding strips is a creative, time-consuming process. Creators have to generate stories from code (code$\mapsto$story) and design comics from stories (story$\mapsto$comic). We contribute CodeToon, a comic authoring tool that facilitates this code-driven storytelling process with two mechanisms: (1) story ideation from code using metaphor and (2) automatic comic generation from the story. We conducted a two-part user study that evaluates the tool and the comics generated by participants to test whether CodeToon facilitates the authoring process and helps generate quality comics. Our results show that CodeToon helps users create accurate, informative, and useful coding strips in a significantly shorter time. Overall, this work contributes methods and design guidelines for code-driven storytelling and opens up opportunities for using art to support computer science education.
\end{abstract}

\begin{teaserfigure}
    \centering
    \includegraphics[trim=0cm 0cm 0cm 0cm, clip=true, width=\textwidth]{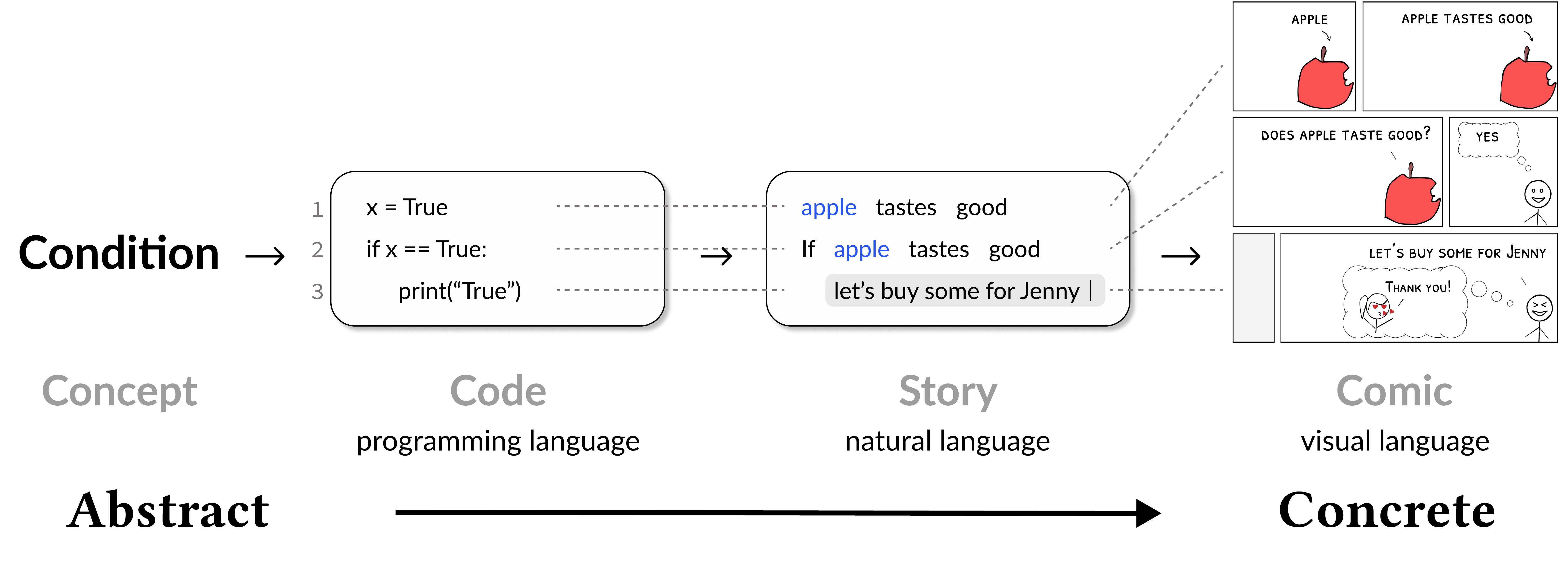}
    \caption{CodeToon helps users create stories and comics from code. It uses 1-to-1 mapping to make connections clear across code, story, and comic. For example, as indicated by the dotted line, line 1 (code) maps to line 1 (story) and to row 1 (comic).}
    \label{fig:abstraction}
\Description[A teaser figure showing the general process taken to generate comics from code. It shows a sequence comprised of 4 stages, starting with the text 'condition' as an example of a programming concept, a piece of code, a story example, and then a comic.]{A teaser figure that shows the general process taken to generate comics from code. The sequence consists of 4 stages, starting with the text 'condition' as an example of a programming concept, a piece of code, a story example, and then a comic. Arrows point to the next stages in this sequence. The code example shown in the sequence is: line 1: x = True; line 2: if x == True; line 3: print(True). The story example that follows this code is: â€œline 1: apple tastes good; line 2: if apple tastes good; line 3: letâ€™s buy some for Jenny. The subsequent comic illustrates this story and is structured as follows: row 1, panel 1: an image of apple with text apple and arrow pointing from the text to the image; row 1, panel 2: the same image of apple and text saying, â€˜apple tastes goodâ€™; row 2, panel 1: the same image of apple and text saying, â€˜does apple taste good?â€™; row 2, panel 2: a stick figure saying, â€˜yesâ€™; row 3, panel 1: a gray empty panel.}
\end{teaserfigure}



\maketitle

\section{Introduction}

\begin{displayquote}
``Computer science is a field that attracts a different kind of thinker\ldots they are individuals who can rapidly change levels of abstraction, simultaneously seeing things `in the large' and `in the small'.''~\cite{hartmanis1994turing}
\end{displayquote}

Learning programming is difficult due to its abstract nature~\cite{suh2020coding}: it requires learning concepts and programming languages that have been derived through a series of abstractions. Specifically, learning with text-based programming languages poses a barrier for novice learners~\cite{hayakawa1947language, suh2020promoting}, as text-based programming languages rely on symbolic representations that use a set of seemingly arbitrary rules and abstract expressions. The compact syntax and notations are useful for specifying precise operations but not for communicating to learners the underlying computational ideas intuitively, unlike pictorial or visual representations, which can leverage real-life scenarios to help learners understand computational concepts. As such, novice learners are often forced to memorize the rules and code expressions without understanding the intuitions behind the syntax and semantics. Unfortunately, this has perpetuated an image of computer programming as a set of abstract ideas and rules, and computer science as an abstruse, inaccessible, and uninteresting discipline, especially for those who struggle with abstract reasoning.

To address this, many researches looked at embodied approaches, exploring ways to use familiar abstractions such as real-life objects, situations, and visual representations to make computer programming more concrete and accessible~\cite{kelleher2007storytelling, resnick2009scratch, guo2013online, suh2020promoting}.
Recent research on \textit{coding strips}, a form of comic strips with code, follows this line of work by looking at comics as a vehicle. By identifying many design variations and patterns for explaining code executions and semantics, Suh~\cite{suh2022phd} showed that comics can be a powerful medium for visualizing computational concepts and procedures.
In another study, Suh et al.~\cite{suh2021using} tested four use cases of coding strips in an introductory computer science course and found that coding strips can enhance learning in various ways. For instance, one use case included an instructor introducing code expressions with comics first and then with code. Students appreciated this scaffolding over the code-only approach as comics allowed them to learn the intuition without being distracted by the syntax and rules and then pick up code expressions in terms of familiar dialogues and actions.

Unfortunately, despite growing evidence of their usefulness, creating coding strips remains a creative, laborious, and time-consuming process.
First, it requires creators to ideate (brainstorm) and select stories that align with code. Second, creators need to invest significant effort and time (and sometimes confidence in drawing) to sketch stories in the form of comics. While Suh et al.~\cite{suh2020coding} proposed a design process and tools to help creators design coding strips, the entire process was manual and not automated~\cite{suh2020coding}. Moreover, while the related literature and previous work suggest that making connections between code and comics obvious is critical for coding strips' success~\cite{ainsworth2006deft, suh2021using}, no work has yet explored how we can establish a clear mapping between code and comics.

We introduce CodeToon, a comic authoring tool that supports this creative process with two mechanisms: (1) facilitating---through metaphor recommendation---the ideation of code-aligned stories and (2) automating the generation of comics from stories. Inspired by Gentner's structure mapping theory~\cite{gentner1983structure, gentner1988metaphor} and visual narrative grammar~\cite{cohn2013visual} for comics, the two mechanisms allow CodeToon users to add code or select code examples provided by the tool, generate a story from the code, and automatically produce comics based on the code or story while maintaining 1-to-1 mapping across them. Our two-part evaluation of CodeToon found that this streamlined design process allows users to quickly and easily create quality coding strips that convey a salient connection between code and comics. In summary, our contributions include:

\begin{itemize}
    \item a computational pipeline that uses story ideation, auto comic generation, and structure mapping for code-driven storytelling;
    \item CodeToon, a tool for code-driven storytelling where users can efficiently transform code into story, then story into comics;
    \item user experiments that evaluate the authoring process and the generated results of CodeToon.
\end{itemize}

\section{Background} \label{related_work}

\subsection{Building Ladder of Abstraction}
\label{structure_mapping_theory}

Coding strip was inspired by the ladder of abstraction, with comic and code representing different levels within the ladder~\cite{codingstrip, suh2020promoting, suh2022phd}.
Thus, we review previous work that addressed two questions for building the ladder of abstraction:
\textbf{Q1}: How do we conceive abstractions at different level(s), and \textbf{Q2}: What are design considerations?

\textbf{Q1$\Rightarrow$A: Find what we can abstract over/under.} In his interactive article \textit{Up and Down the Ladder of Abstraction}~\cite{victor2011}, Bret Victor uses variable as a control for moving up and down the abstraction ladder. 
In this article, the variable is \code{time}, and a system at a particular \code{time} an abstraction; readers use a slider to change the \code{time} (e.g., \code{t=1} to \code{t=2}) and observe how the system (abstraction)---a car's trajectory---changes. He equates this interaction as moving up and down the ladder of abstraction, explaining that all systems share the same anatomy---an independent variable (e.g., \code{time}), structure (the set of rules and what is controlled by the variable), and data (environment)---and suggesting that the process of building the ladder of abstraction (i.e., conceiving abstractions at different levels) consists of identifying what can be parameterized and providing a control to explore the range of abstractions. This informed how we should engineer story ideation (code$\mapsto$story). Specifically, in conceptualizing how we can turn code into a story, we used this idea to explore what parts of the code can be parameterized and used to develop stories. (Further details in Section~\ref{sec:design-choices}.)

\textbf{Q2$\Rightarrow$A: Maintain structure across abstractions.} Defining abstraction as ``a comparison in which the base domain [(e.g., code)] is an abstract relational structure,'' Gentner proposed structure mapping theory to posit that the structure---the relations between objects---is the most important factor when conceiving new abstractions, not the number of attributes shared between the base and target domains or the specific content~\cite{gentner1988metaphor}. An example he gives is: ``The hydrogen atom is like our solar system.'' In this example, what makes the hydrogen atom a comparable abstraction is that the hydrogen atom and solar system share the same relation (e.g., the electron REVOLVES around the nucleus, like how the planets REVOLVE around the sun), not their object attributes (they do not share the same object attributes, e.g., color, size, as the planets). A related technique called concreteness fading, which introduces an idea in multiple stages using different representations (abstractions) in decreasing concreteness, also supports this, suggesting that maintaining the relational structure across the representations is the key~\cite{fyfe2018making, suh2019using, suh2020we}. This design principle for layers of abstraction inspired us to structure stories and comics to align with the code structure.

\subsection{Supporting Comic Authoring}

\subsubsection{Design Process \& Patterns} The time-consuming, laborious nature of creating comics poses critical barriers to their use and adoption. As a result, various approaches have been suggested. For example, to support ideation in the design process, researchers developed design patterns and process with clearly delineated stages to guide authors~\cite{bach2018design, suh2020coding}. Digital authoring tools have been developed to make it easier to quickly draft and iterate on the design by offering templates (e.g., panel layout and images) users can add to the canvas~\cite{pixton, kim2019datatoon, suh2022privacytoon}. Our work extends prior work as our tool supports the authoring of comics (1) based on code input and (2) introduces two new mechanisms---story ideation and auto comic generation---on top of the methods mentioned above.

\subsubsection{Automatic Comic Generation}\label{auto-comic-generation}

\begin{figure}[tb!]
    \centering
    \includegraphics[trim=0cm 0cm 0cm 0cm, clip=true, width=0.48\textwidth]{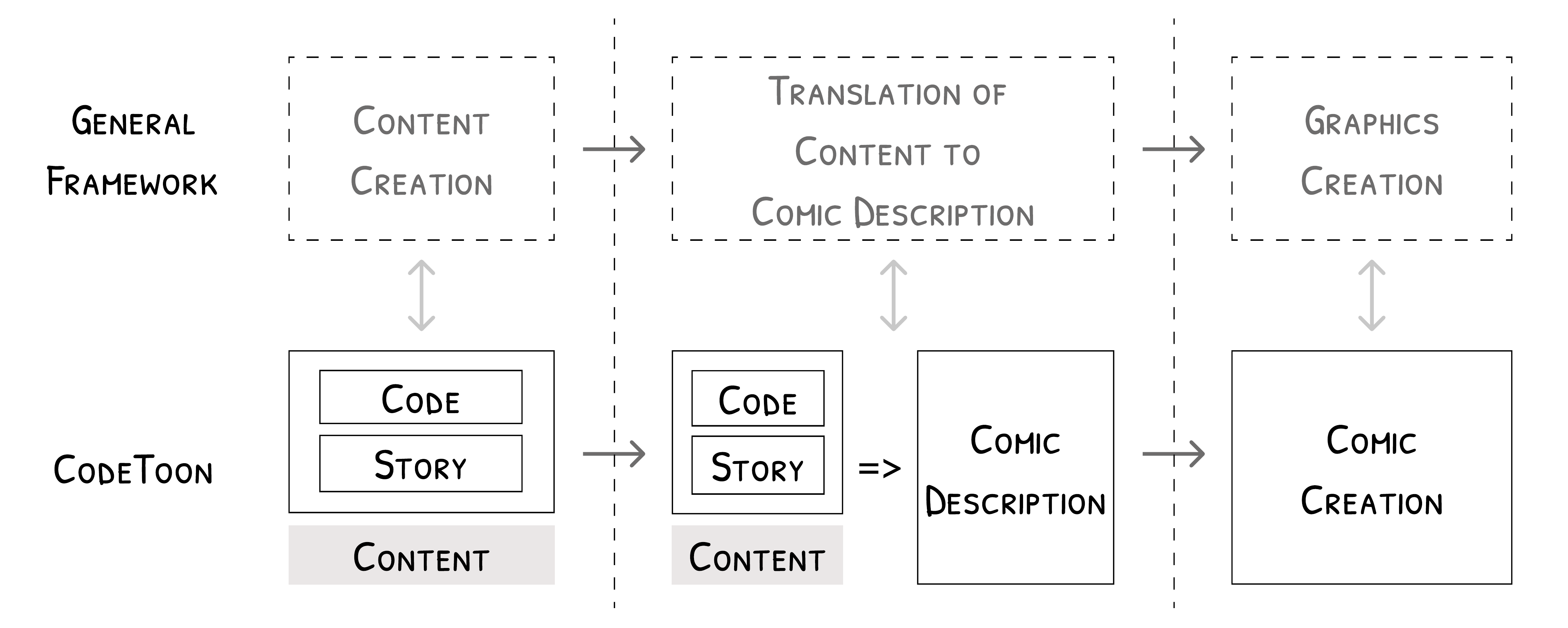}
    \caption{General framework for auto generation of comics suggested by Zeeders~\cite{zeeders2010comics} and our framework. In our framework, content used to generate comics are code and story. CodeToon users add content (code \& story), and the system translates it into comic description and creates a comic strip, the graphic representation of code \& story.}
    \label{fig:autocomic_generation}
\Description[A general framework for auto-generation of comics suggested by Zeeders. It also shows how CodeToonâ€™s auto comic generation process fits in relation to it.]{A general framework for auto-generation of comics suggested by Zeeders. It also shows how CodeToonâ€™s auto comic generation process fits in relation to it. The general framework is a simple flowchart that proceeds from (1) content creation to (2) translation of content to comic creation, and to (3) graphics creation. At a high-level, CodeToon also consists of these three stages. The figure shows code and story as content for the first stage (content creation); the second stage also shows code and story being converted to comic description; the third stage is shown as comic creation.}
\end{figure}

One step forward from supporting a design process with authoring tools, design patterns, and design guidance is automatically generating comics for users~\cite{cao2012automatic, kang2021toonnote, wang2021interactive}. Zeeders~\cite{zeeders2010comics}---who surveyed auto comic generation methods---suggests that, at a high-level, auto comic generation involves three steps: (1) content creation, (2) translation of content to a comics description, and (3) graphics creation, as shown in Fig.~\ref{fig:autocomic_generation}. In prior work, the sources of content in the first step have been a multitude of things, e.g., daily activity data~\cite{cho2007generating}, chat sessions~\cite{kurlander1996comic}, scripts~\cite{kesiev2011}, and movies~\cite{hong2010movie2comics, yang2021automatic}. (Their work cannot generalize to our content type, code, since it, unlike other data types, lacks contextual information that can be used to form a narrative for a comic without the user intervening to help define a story.)
In the second step, they are formatted into a particular format~\cite{alves2007comics2d, zeeders2010comics} to provide instructions on how they should be presented graphically. The final step is the graphics creation stage, where either the composition or screenshot method is used~\cite{zeeders2010comics}; the former method composites different images (e.g., character, background, speech bubble) to create a scene for panels, the latter embeds existing scenes (e.g., screenshot of movie scenes) into panels. 
As we will explain, our work leverages the composition method to generate comics automatically. Overall, our work extends research in this area by demonstrating how we can auto generate comics from a new content type, code.

\section{CodeToon}\label{codetoon-section}

\subsection{Design Goals}\label{design_goals}

To develop CodeToon, we first conducted a pilot study with 12 participants.
Two participants (age: M=44.5; gender: 1F, 1M) were teachers with 5+ years of experience teaching programming; ten other participants (age: M=27.9; gender: 5F, 5M) were undergraduate and graduate students with varying teaching experience (6 0-1 year, 2 1-3 years, 1 5+ years).
We chose teacher and student participants highly experienced in programming (11 Much Experience, 1 Some Experience) as opposed to participants without programming experience in order to harness the insights they picked up over the years as teachers and students.
Over the course of the pilot sessions, we improved, added, and tested new features and workflows of CodeToon (Sections~\ref{user_interface} and \ref{sec:design-choices}) until we could observe that no major changes are needed to enable a creative authoring experience.  
Based on this pilot study, the literature on multiple representational systems~\cite{ainsworth2006deft, berthold2009instructional, suh2020we}, and creativity support for comics~\cite{suh2020coding}, we derived the following design goals for CodeToon:

\begin{figure*}[htb]
    \centering
    \includegraphics[trim=0cm 0.5cm 0cm 0cm, clip=true, width=\textwidth]{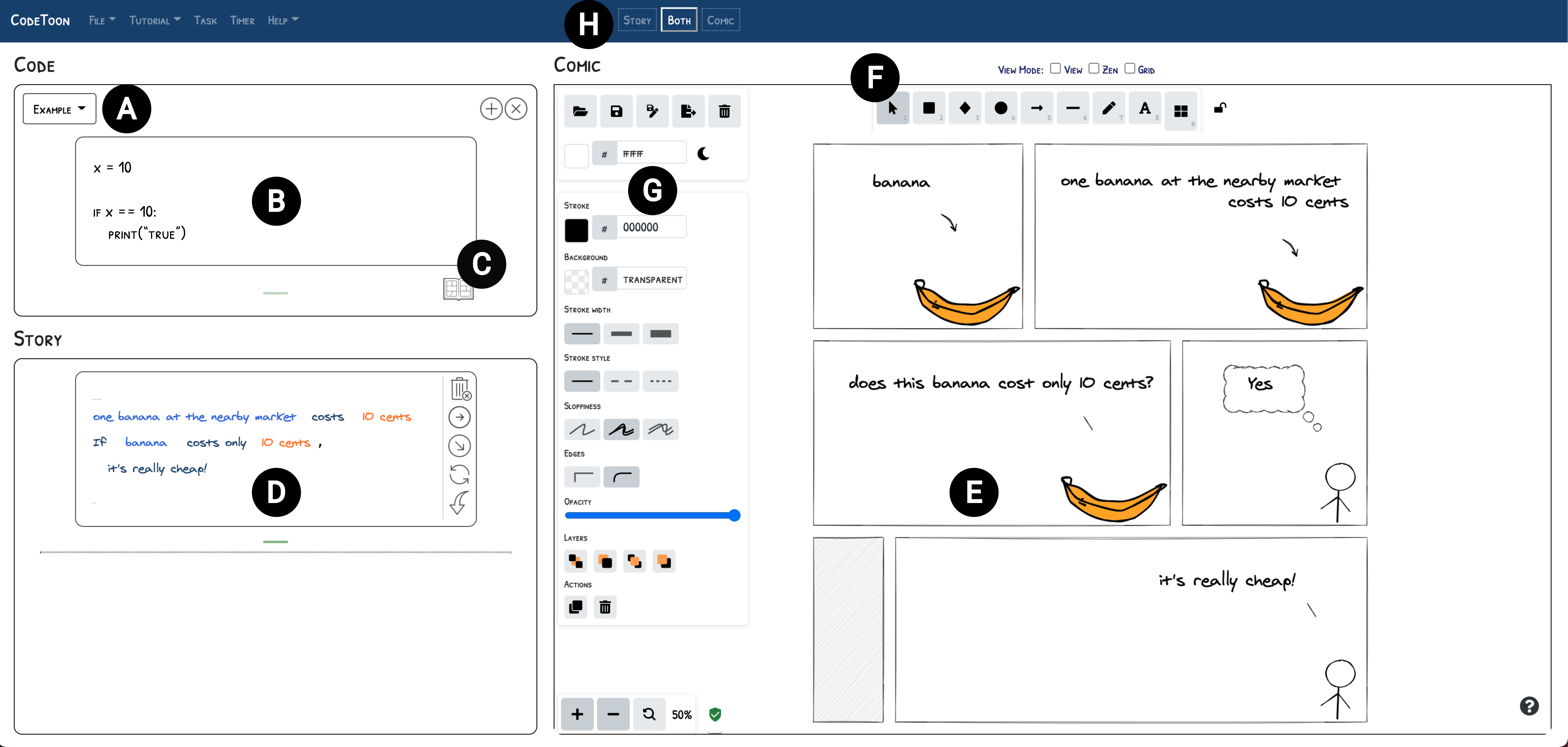}
    \caption{System interface: (A) drop-down allows users to check potential code examples for basic programming concepts, (B) code container, (C) button that generates story template from code in code container, (D) story template, (E) drawing canvas for comics, (F) tool palette, (G) style palette, and (H) buttons for changing the interface layout (between code \& story, current, or canvas-only layout).}
\Description[A screenshot of the system interface that consists of three panels: code, story, and comic.]{A screenshot of the system interface. The interface consists of three panels: code, story, and comic. The panels contain examples of code, story, and comic. The code panel contains a code container for users to add code; the story panel contains a story template generated based on the code; the comic panel shows a canvas with tool palette and the canvas also features a comic auto generated from the story template.}
  \label{fig:interface}
\end{figure*}

\textbf{D1. Allow users to iterate on their code, story, and comic}.
From our pilot study, we found that the authoring process may not be linear. While creating comics, users can be inspired by their comics and form additional ideas to add to their story. While working on the story template, they could also think of a better story and desire to edit code (e.g., change the value assigned to a variable). Thus, the tool should make this interaction easy for its users.
    
\textbf{D2. Augment, not constrain, users' creativity with our story ideation and auto comic generation.}\label{d2} 
Our pilot study revealed that providing story ideas and comic templates can accelerate the authoring process. But it also showed that some users can already have some ideas on what they want to create and how to design their comics. Thus, the tool should not limit users to using only story ideas and comic templates provided by the tool.

\textbf{D3. Make mapping clear across code, story, and comic}.
Research suggests that making the correspondence explicit and consistent is essential when presenting multiple representations~\cite{suh2020we, suh2021using}. Otherwise, they do more harm than good because they only confuse people. For us, this means that the mapping ($\leftrightarrow$) between code, story, and comic should be clear. Previous research on coding strip also found that the mapping between code and comics needs to be clear for it to be effective and useful~\cite{suh2021using}.

\textbf{D4. Use simple, scalable visual vocabulary.}
Scalability relies on having a set of basic building blocks that can be combined to build anything of varying complexity (cf. the composition method in Section~\ref{auto-comic-generation}). A set of building blocks in visualizations is called visual vocabulary. To generate comics that can scale to any code input, establishing a simple, scalable set of comic templates that can be easily combined to express any set of code is necessary.

As a whole, our design goals aimed to create a `low floor, high ceiling' system for generating coding strips. That is, a system that makes the process of creating coding strips simple, effortless, and easy, while providing a high ceiling for creative exploration.

\subsection{User Interface}\label{user_interface}

CodeToon consists of three panels: code (Fig.~\ref{fig:interface}B), story (Fig.~\ref{fig:interface}D), and comic (Fig.~\ref{fig:interface}E). Users can select any layout button (Fig.~\ref{fig:interface}H) to change which panels are shown. The default layout (\textsc{Both}) shows all three panels (Fig.~\ref{fig:interface}), the \textsc{Story} layout code and story panels, and the \textsc{Comic} layout the drawing canvas. 
The layout feature allows users to customize the workspace---the number and layout of the panels---for an optimal experience.

The three panels represent stages in the design process for creating a coding strip~\cite{suh2020coding}. Concretely, the basic workflow consists of users (1) adding code in the code panel (Fig.~\ref{fig:interface}B), (2) generating a story template (from the code) and writing a story in the story panel (Fig.~\ref{fig:interface}D), and (3) using the auto comic generation feature to instantly generate comic in the comic panel (Fig.~\ref{fig:interface}E). Below, we describe each panel and how they facilitate this workflow.

\begin{figure}[tb]
    \centering
    \includegraphics[trim=0cm 0cm 0cm 0cm, clip=true, width=0.48\textwidth]{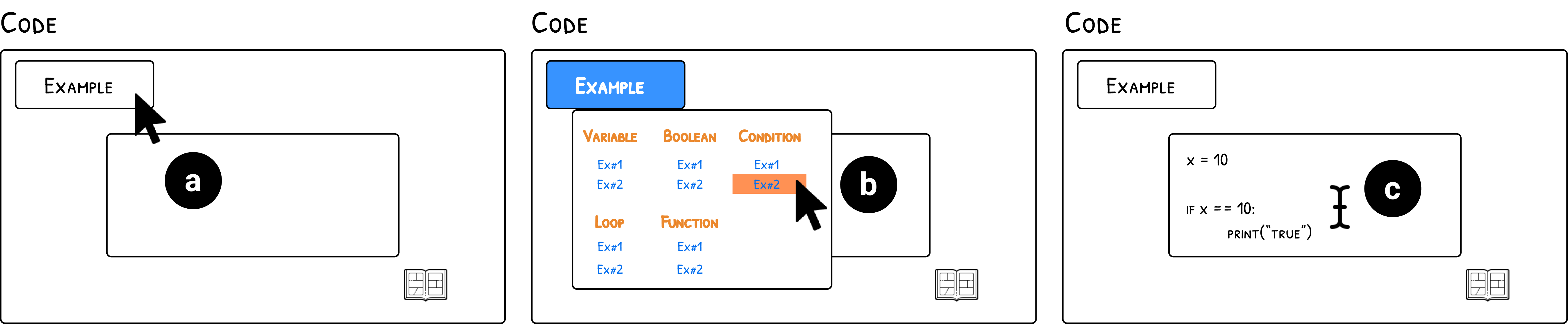}
    \caption{Users can add code by selecting code example (b) or by typing (c).}
    \label{fig:workflow_ac}
\Description[A figure showing how users can add code to a code container by selecting a code example from a dropdown or by typing.]{It shows how users can add code to a code container by selecting a code example from a dropdown or by typing.}
\end{figure}

\subsubsection{Code.}\label{code-section} In the code panel (Fig.~\ref{fig:interface}B), users can add any number of programs, each within a different code container.
As shown in Fig.~\ref{fig:workflow_ac}, users can add the code to the container by using the code example repository (Fig.~\ref{fig:interface}A) or manually typing into the code container. There are two buttons in the top right corner of the code panel for adding (\,\raisebox{-2pt}{\includegraphics[scale=0.1]{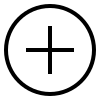}}\,) and deleting (\,\raisebox{-2pt}{\includegraphics[scale=0.1]{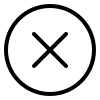}}\,) code containers. The ability to add additional code containers was added during the pilot phase to make it easy for users to iterate on their code, story, and comic (\textbf{D1}). After the user adds code, they can press a button (Fig.~\ref{fig:interface}C,\,\raisebox{-2pt}{\includegraphics[scale=0.1]{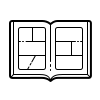}}\,) to generate a story template (Fig.~\ref{fig:interface}D) from code.

\begin{figure}[tb]
    \centering
    \includegraphics[trim=0cm 0cm 0cm 0cm, clip=true, width=0.48\textwidth]{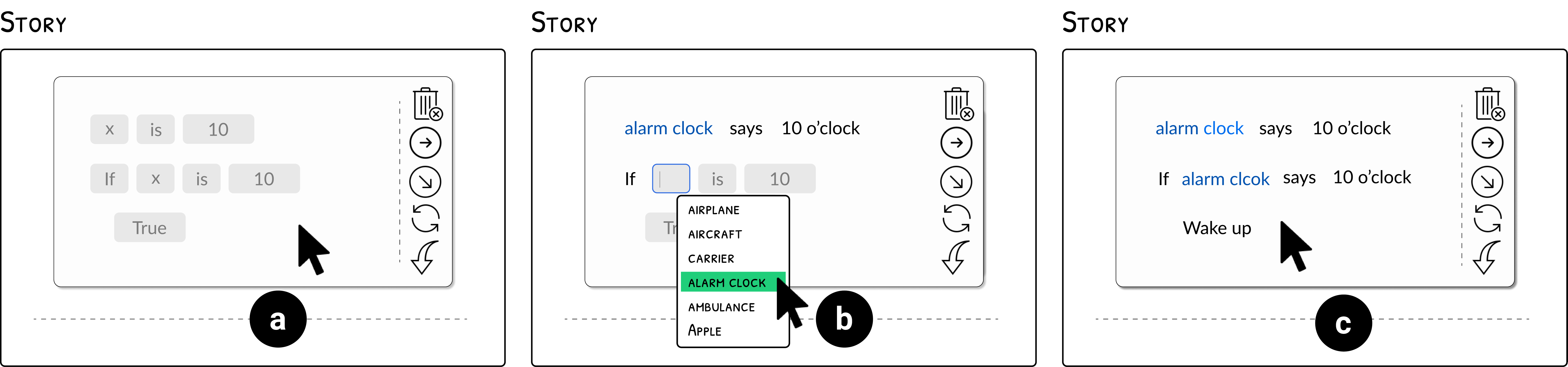}
    \caption{Users can add story (to story template) by selecting a list of ideas in dropdown (b) or by typing into input box.}
    \label{fig:workflow_as}
\Description[A figure showing how users can add a story to a story template by selecting from a list of ideas in the dropdown or by typing into the template.]{It shows how users can add a story to a story template by selecting from a list of ideas in the dropdown or by typing into the template.}
\end{figure}

\subsubsection{Story.}\label{story-section} When a user generates a story template, it is added to the story panel, which is initially an empty panel. Fig.~\ref{fig:interface}D shows what the user would see when a story template is added. Story templates are linguistic representation of the code, with input boxes where users can add real-life equivalents for the code expressions. For instance, the code expression \code{x = 10} generates \,\raisebox{-2pt}{\includegraphics[scale=0.1]{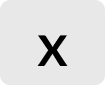}} \,\raisebox{-2pt}{\includegraphics[scale=0.1]{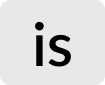}} \,\raisebox{-2pt}{\includegraphics[scale=0.1]{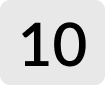}} (Fig.~\ref{fig:workflow_as}(a-b)) as its story template. Now, a user can add \texttt{alarm clock} to \,\raisebox{-2pt}{\includegraphics[scale=0.1]{figures/icon_x.png}}, \texttt{says} to \,\raisebox{-2pt}{\includegraphics[scale=0.1]{figures/icon_is.png}}, and \texttt{10 o'clock} to \,\raisebox{-2pt}{\includegraphics[scale=0.1]{figures/icon_10.png}}. As shown in Fig.~\ref{fig:workflow_as}(b), CodeToon provides a dropdown containing a list of metaphors to help users brainstorm story ideas. The dropdown does not appear for every input box, however. At the time of testing, it appeared only on input boxes mapped to variables (e.g., \code{x}) and assignment operators (\code{=}). The dropdown for the former showed a list of 345 categories (e.g., apple, car) and that of verbs synonymous with or semantically close or related with the semantics of assignment operator (e.g., assign, has) for the latter. While dropdowns appear to help users with story ideation, they do not have to form their stories around these suggestions; they can type any text into the input box to create any story (\textbf{D2}).

\begin{figure}[h]
    \centering
    \includegraphics[trim=0cm 0cm 0cm 0cm, clip=true, width=0.48\textwidth]{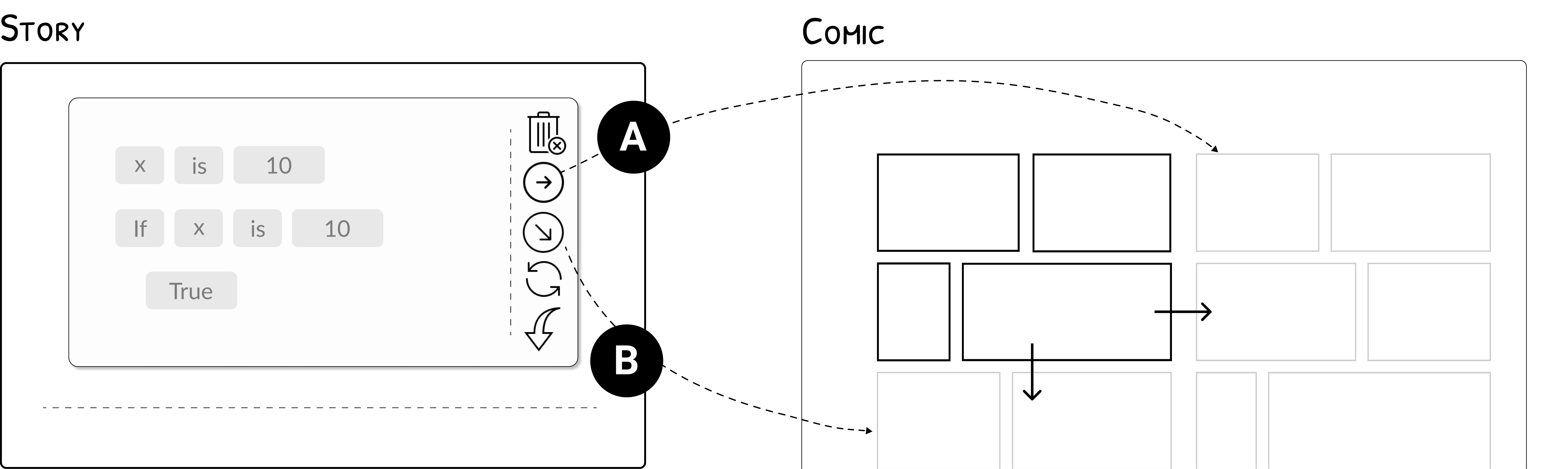}
    \caption{Users can generate a comic by selecting either of the arrow buttons. If canvas already has some drawing elements, as in this case, they can add the comic to the right (A) or below (B) the existing elements. If canvas is empty, both buttons place the comic at the center of the canvas.}
\Description[A figure showing how users can auto generate a comic from the story template.]{It shows how users can auto generate a comic from the story template. It shows two buttons in the story template, which auto generate comics and add to the canvas in the comic panel. It also indicates that users can expand the existing comic. To explain, the two buttons are there to allow users to specify whether to add a comic to the next or below the existing, if any, elements in the canvas.}
    \label{fig:workflow_gc}
\end{figure}

\subsubsection{Comic.}\label{comic-section} 
\begin{figure}[th]
    \centering
    \includegraphics[trim=0cm 0cm 0cm 0cm, clip=true, width=0.35\textwidth]{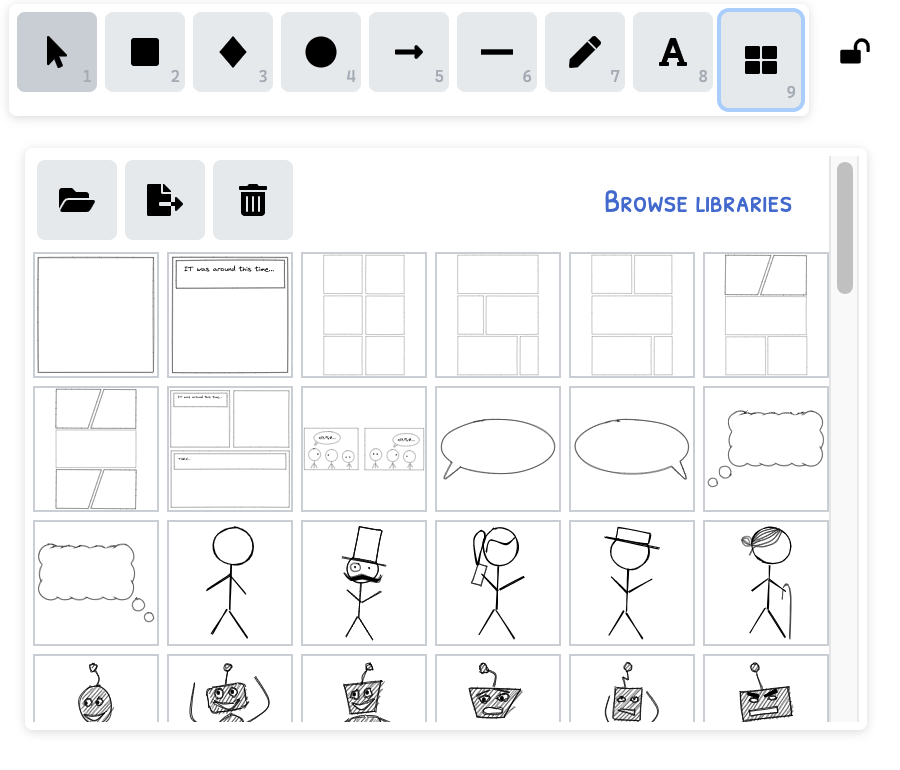}
    \caption{The library offers pre-drawn templates, such as panels, speech bubbles, and characters, for creating comics.}
    \label{fig:library}
\Description[A screenshot of the template library that users can use to add pre-defined templates, such as comic panels, speech bubbles, and characters.]{It shows a screenshot of the template library that users can use to add pre-defined templates, such as comic panels, speech bubbles, and characters.}
\end{figure}

As shown in Fig.~\ref{fig:workflow_gc}, a user can instantly generate a comic (Fig.~\ref{fig:interface}E) by selecting any of the two arrow icons (\,\raisebox{-2pt}{\includegraphics[scale=0.1]{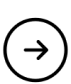}} and \raisebox{-2pt}{\includegraphics[scale=0.1]{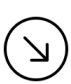}}\,) below the trash can icon \,\raisebox{-2pt}{\includegraphics[scale=0.1]{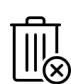}}.
The reason for the two arrows is to offer users the flexibility to expand to the right or below the existing drawings.
If users change the story, they can press the update icon \,\raisebox{-2pt}{\includegraphics[scale=0.1]{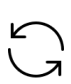}} (below the arrow icons) to instantly update the content of the auto generated comic to reflect these changes, making iterative design of their story and comic frictionless (\textbf{D1}).
While users can use the auto comic generation feature to instantly generate comics from the story template, they can also use the tool palette (Fig.~\ref{fig:interface}F), style palette (Fig.~\ref{fig:interface}G), and library (Fig.~\ref{fig:library}) to manually create or edit/expand on the auto generated comic.
The view mode above the palette (Fig.~\ref{fig:interface}F) provides three checkboxes for turning on/off different view modes: \textsc{Grid} makes the grid appear in the canvas for precise alignment and measurement, \textsc{Zen} removes style palette, and \textsc{View} removes both the tool and style palette and makes the canvas view only.

\subsection{Usage Scenarios}\label{workflow}

To further clarify the user interface and workflow, we present two scenarios to demonstrate how users can use CodeToon. 

\textbf{Teacher.} Amanda is a high school teacher. She is teaching a programming class to 10th graders who are learning programming for the first time. She wants her students to discover that programming is not just about memorizing rules, syntax, and expressions. She wants her students to realize that computational ideas can also be found in our daily lives. When the class starts, she explains this as her goal for her students. To show how they can think of computing in terms of real-life objects and situations, she opens CodeToon and shows a sequence of the code-story-comic example shown in Fig.~\ref{fig:abstraction}. She explains that while code expressions may appear scary for now, they can think of code as being no different from words we use to communicate. To direct their attention to the logic (i.e., \textsc{if-then} structure) and not the content, she switches the object in the story from \,\raisebox{-2pt}{\includegraphics[scale=0.1]{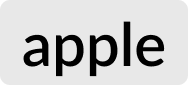}}  to  \raisebox{-2pt}{\includegraphics[scale=0.1]{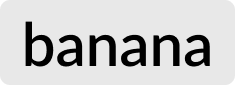}\, }
  and updates the comic (by pressing the update icon \,\raisebox{-2pt}{\includegraphics[scale=0.1]{figures/icon_refresh.png}}\,). To make the stories more engaging and help students connect with them, she also adds contextual details to the expression \,\raisebox{-2pt}{\includegraphics[scale=0.1]{figures/icon_banana.png}}, editing it to \,\raisebox{-2pt}{\includegraphics[scale=0.1]{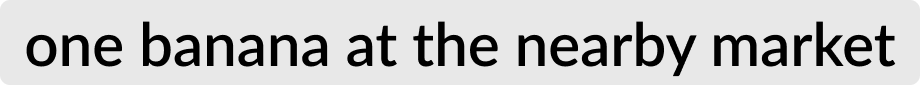}}, generating a sentence \,\raisebox{-2pt}{\includegraphics[scale=0.1]{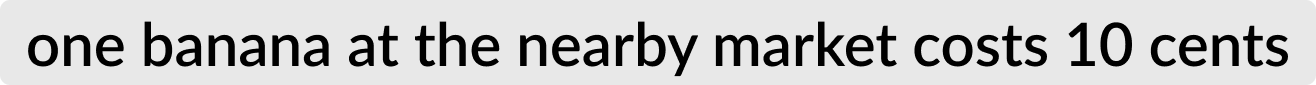}} (cf. Fig.~\ref{fig:interface}). She invites her students to suggest a story and produces comics for them on the fly. Students suggest diverse stories based on their experience and learn that programming is not as difficult as they thought it would be.

\begin{figure}[h!]
    \centering
    \includegraphics[trim=0cm 0cm 0cm 0cm, clip=true, width=0.48\textwidth]{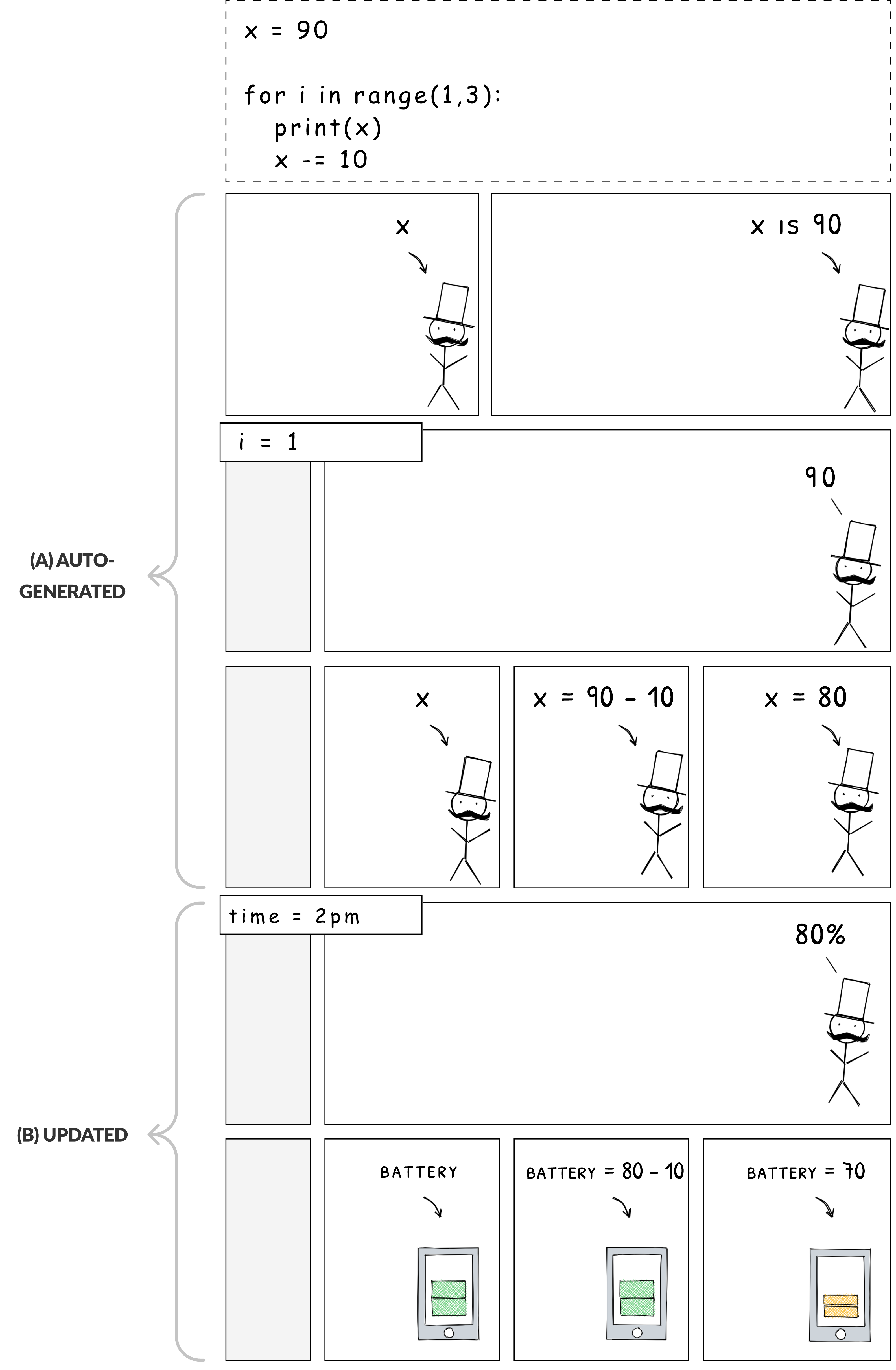}
    \caption{A simple loop code and (A) auto generated comic. The bottom two rows represent (B) comic updated with different images (\textsf{phone}) and text (e.g., \textsf{BATTERY} for \code{x}). Like in this example, CodeToon can update specific rows, giving users fine-grained control over their stories.}
\Description[An example of an auto generated comic based on a loop code.]{It shows an example of an auto generated comic based on a loop code. The comic spans 5 rows of panels, with the first three rows representing auto generated comics and the bottom two rows updated with different images and text. Whereas the auto generated comic featured a stick figure, the replaced image showed a phone; text was also updated from, for instance, x to battery.}
    \label{fig:user-scenario}
\end{figure}

\textbf{Student.} Jane is a graduate student working as a teaching assistant in the introductory CS course. While overseeing a lab, a student asks for help, saying his code is not working the way he wants it to. She takes a look at the code and feels that a piece of code involving loop may be the culprit. She tells the student but realizes that he does not have a good grip on the concept of loop and how to debug the code. She opens CodeToon and adds the code in his assignment and generates a comic to show the student how loop works. Fig.~\ref{fig:user-scenario} shows an example of a code with loop and its corresponding comic. Jane explains to the student how the first row of the comic maps to the first line of the code (\code{x = 90}) and that the next two rows represents the two lines in the loop; She notes that the program then moves to the next iteration and points to \code{i = 1} as an indicator of which iteration the program is running at. To help the student connect the concept to a real-life situation, Jane adds a story to the story template: she replaces \,\raisebox{-2pt}{\includegraphics[scale=0.1]{figures/icon_x.png}}  with \,\raisebox{-2pt}{\includegraphics[scale=0.1]{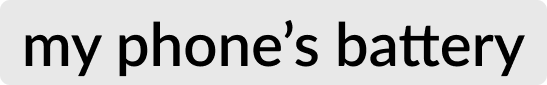}}, \,\raisebox{-2pt}{\includegraphics[scale=0.1]{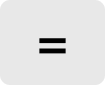}} with \,\raisebox{-2pt}{\includegraphics[scale=0.1]{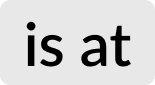}}, \,\raisebox{-2pt}{\includegraphics[scale=0.1]{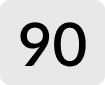}} with \,\raisebox{-2pt}{\includegraphics[scale=0.1]{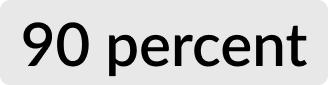}}, and \,\raisebox{-2pt}{\includegraphics[scale=0.1]{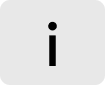}}  with \,\raisebox{-2pt}{\includegraphics[scale=0.1]{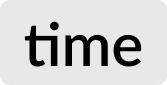}} and presses the update icon to update the comic, which shows a comic with a story that shows the phone having a battery initially at 90\% and decreasing by 10\% every hour, as shown in Fig.~\ref{fig:user-scenario}. Jane creates another code container and shows the student another loop example to help him master the concept. After the student is done receiving help from Jane, he asks Jane for the URL of CodeToon so that he can use it to review and assist his studies in the class.

\section{Code-Driven Storytelling} \label{sec:design-choices}

CodeToon supports code-driven storytelling by generating comics from code using two mechanisms---story ideation and auto comic generation. In this section, we describe how we designed them. The implementation details of the computational pipeline are described in Fig.~\ref{fig:baseline-effort-examples} in Appendix.

\subsection{Story Ideation}

\begin{figure}[h]
    \centering
    \includegraphics[trim=0cm 0.5cm 0cm 0cm, clip=true, width=0.48\textwidth]{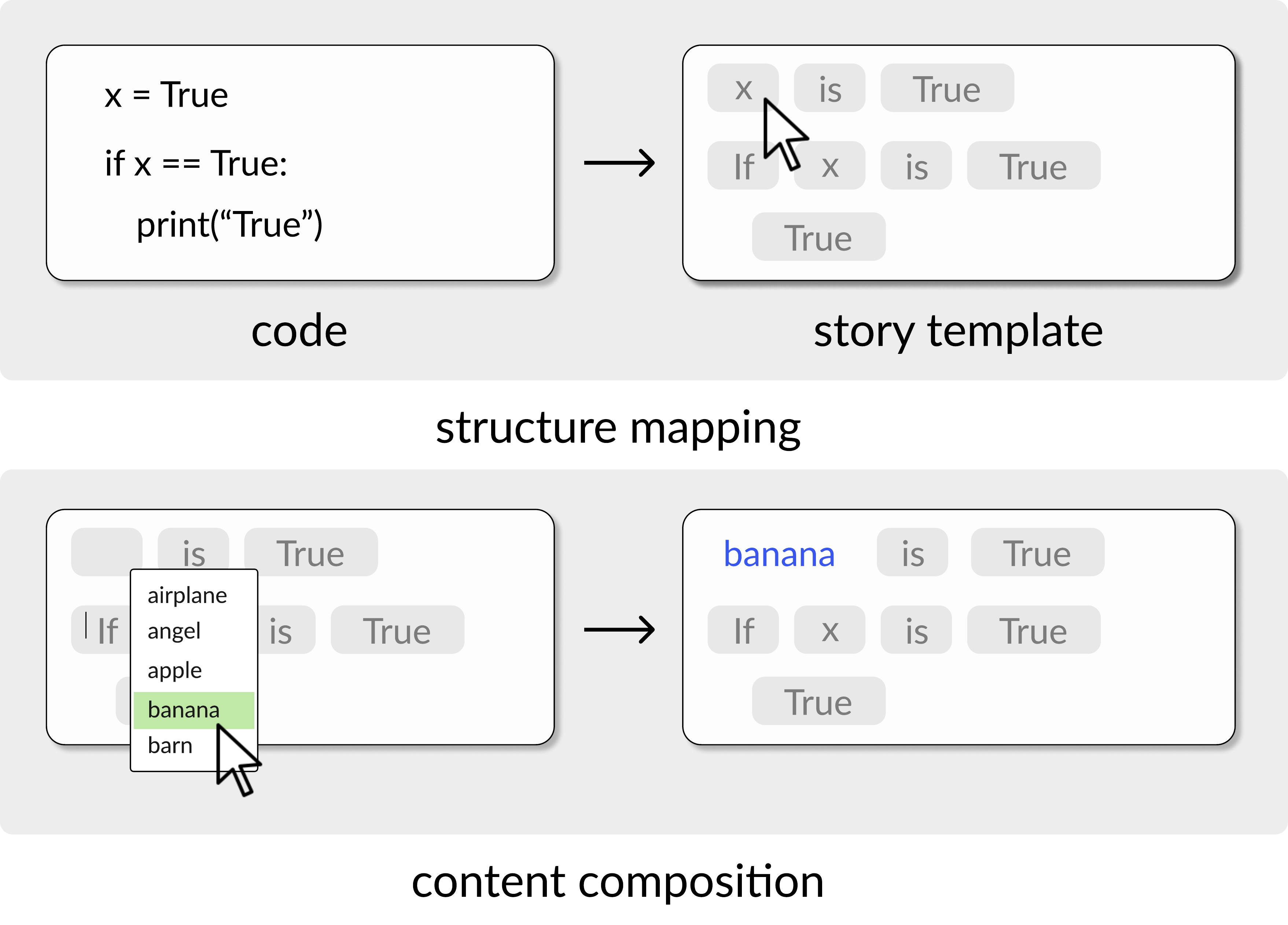}
    \caption{Our story ideation consists of generating a story template aligned with the code structure (structure mapping) and allowing users to fill the template with the help of metaphor suggestions (content composition).}
\Description[A figure showing the steps in our story ideation process.]{It shows our story ideation process, which consists of 2 steps: the first step is called structure mapping and shows code being mapped to a story template; the second step is called content composition and shows a user adding a story into the story template using a list of metaphors in the dropdown.}
    \label{fig:structure_mapping}
\end{figure}

CodeToon aims to make the generation of comics from code more efficient, and the first step in that process is rapid story ideation. To understand how we can turn code into a story, we first went through code expressions and turned them into a story. Table~\ref{tab:code-to-story} shows some examples. We found that any code expression can be parameterized and replaced with metaphors of corresponding form (cf. Section~\ref{structure_mapping_theory}). For instance, variables (e.g., \code{x}) can be metaphors in noun (e.g., \textsf{wallet}) or phrases (e.g., \textsf{message in my email}); assignment operator (\code{=}) can be \textit{be} verbs (e.g., \textsf{am}, \textsf{is}, \textsf{are}) and \textit{transitive \& intransitive} verbs (e.g., \textsf{wallet \textit{has} 5 parking coins}, \textsf{my dog \textit{feels} sick}); values can be contextualized (e.g., \code{5}$\rightarrow$\textsf{5 o'clock}, \code{True/False}$\rightarrow$\textsf{on/off}, \code{``hello''}$\rightarrow$\textsf{``hello, John''}); keywords (e.g., \code{def}) and (built-in/user-defined) function names (e.g., \code{print()}) can be replaced with semantically related verbs/phrases. For instance, \code{print} can be \textsf{say}. 
Therefore, in the story template, we (1) provided a list of metaphors they can choose from and (2) converted code expressions into text fields, to allow users to be creative with the story authoring (\textbf{D2}).

\begin{table}[h]
\caption{Examples of code and corresponding stories}
\label{tab:code-to-story}
\centering  
\resizebox{\columnwidth}{!}{
    \centering
    \begin{tabular}{l l l}
    \toprule
         \code{code} & \textit{hybrid} (\code{code} \& \textsf{story}) & \textsf{story} \\
         \midrule 
         \midrule 
         \multirow{3}{*}{\code{x = 5}} & \textit{time = 5} & \textsf{time is 5 o'clock} \\
         & \textit{wallet = 5} & \textsf{wallet has 5 parking coins}\\
         & \textit{student = 5} & \textsf{student received 5 dollars}\\
         \midrule
         \multirow{3}{*}{\code{x = True}} & \textit{switch = on} & \textsf{switch is on} \\
         & \textit{my\_schedule = busy} & \textsf{my schedule is busy} \\
         & \textit{this = True} & \textsf{this is expensive} \\
         \midrule 
         \code{x = ``hello''} & \textit{message = ``hello''} & \textsf{message reads, ``hello''} \\
         \midrule 
         \code{print(``Even'')} & \textit{print(``it's even'')} & \textsf{say, ``It's even!''} \\
         \midrule
    \bottomrule
    \end{tabular}
}
\Description[Examples of code and corresponding stories.]{It shows examples of code and corresponding stories. The table has three columns: code, hybrid (code & story), and story. There are five rows showing examples of code, hybrid, and story. One example of the row is as follows: code: x = 5; hybrid: time = 5; story: time is 5 oâ€™clock.}
\end{table}

\subsection{Auto Comic Generation} 

Auto comic generation requires the conversion of different types of code expressions into a visual panel arrangement (e.g., panels, characters, speech bubbles).  We approach this problem by defining in advance a specific design template for each code expression (e.g., variable assignment, loop).  Specifically, we leveraged the theory of Visual Narrative Grammar (VNG)~\cite{cohn2015analyze}, which suggests that each panel of a comic can be categorized into one of the five phases of a narrative: (1) \textsc{Establisher}, which sets up a scene; (2) \textsc{Initial}, which depicts the start of an action; (3) \textsc{Prolongation}, which shows moments between the start of an action and its peak; (4) \textsc{Peak}, which marks the point in the action when the tension reaches the peak; (5) \textsc{Release}, which shows moments after the action has ended. Below, we provide two examples of code expression template, and the thought process that went into designing these templates using VNG.

\begin{figure}[h]
    \centering
    \includegraphics[trim=0cm 0cm 0cm 0cm, clip=true, width=0.48\textwidth]{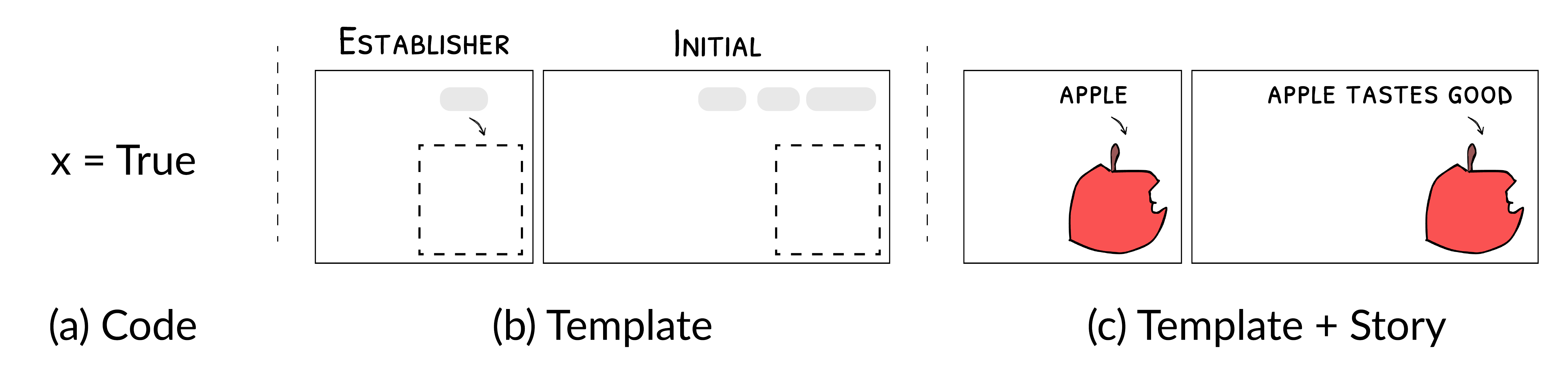}
    \caption{Variable assignment: code, template, example}
\Description[An example of variable assignment code, its comic template, and the template with a story embedded in it.]{It shows an example of variable assignment code, its comic template, and the template with a story embedded in it. Specifically, they are: code: x = true; template: 2 panels labeled, each labeled as establisher and initial; template + story: 2 panels with apple sketches and text, apple and apple tastes good in the two panels, respectively.}
    \label{fig:variable_assignment}
\end{figure}

Fig.~\ref{fig:variable_assignment} shows the variable assignment \code{x = True} (Fig.~\ref{fig:variable_assignment}(a)), its template (Fig.~\ref{fig:variable_assignment}(b)), and the first row of the comic in Fig.~\ref{fig:abstraction} (Fig.~\ref{fig:variable_assignment}(c)) that uses this template. Variable assignment can be defined by two operations: first, computers allocate a memory space for a variable; then they assign value to the variable. The first operation is (semantically speaking) analogous to \textsc{Establisher} in that it \textit{sets up a scene} (for assigning value). The second step can be considered as \textsc{Initial} as it \textit{initiates} the action of assigning value to this variable. Hence, we use two panels, \textsc{Establisher} and \textsc{Initial}, as shown in Fig.~\ref{fig:variable_assignment}. While it can also be valid to have just one panel (e.g., \textsc{Initial} instead of \textsc{Establisher} + \textsc{Initial}), one consideration that led us to choose the \textsc{Establisher} + \textsc{Initial} combination is design. Typical computer programs will contain multiple variable assignments, which means that the auto generated comic will have several templates like this, stacked on top of each other. Having two or more panels in each row, aligned and stacked on top of each other, can make the final comic design look more structured (cf. Fig.~\ref{fig:abstraction}) than the design where multiple rows have only one panel in each row. Additionally, this design can be more useful in the classroom. When teaching variable assignment, for example, having illustrations for space allocation and value assignment can allow teachers to teach what happens at the memory level.

\begin{figure}[h]
    \centering
    \includegraphics[trim=0cm 0cm 0cm 0cm, clip=true, width=0.48\textwidth]{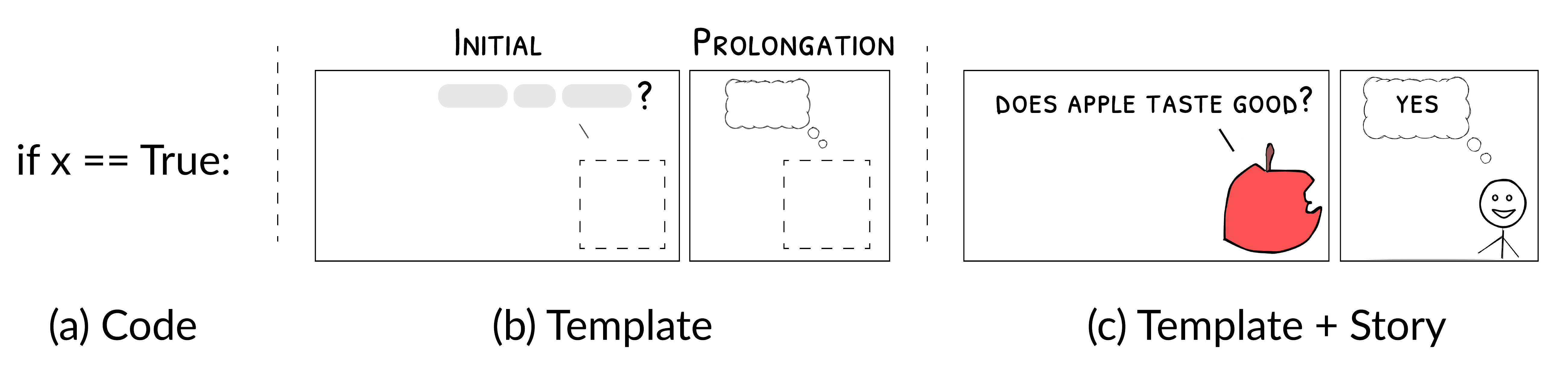}
    \caption{Conditional expression: code, template, example}
    \label{fig:conditional}
\Description[An example of conditional code, its comic template, and the template with a story embedded in it.]{It shows an example of conditional code, its comic template, and the template with a story embedded in it. Specifically, they are: code: if x == true; template: 2 panels labeled, each labeled as initial and prolongation; template + story: 2 panels with an apple and stick figure and text, does apple taste good? and yes in the two panels, respectively.}
\end{figure}

As another example, Fig.~\ref{fig:conditional} shows the conditional expression \code{if x == True:} (Fig.~\ref{fig:conditional}(a)), its template (Fig.~\ref{fig:conditional}(b)), and the second row of the comic in Fig.~\ref{fig:abstraction} (Fig.~\ref{fig:variable_assignment}(c)) that uses this template. A similar thought process went into designing this template: the conditional expression first checks whether the statement is \code{True} or \code{False}; as this \textit{initiates} an action, the first panel is an \textsc{Initial} panel and has a placeholder for text that ends with a question mark to indicate the checking action. The following panel is used to report (hence the speech bubble) whether the expression evaluates to \code{True} or \code{False}; this panel is a \textsc{Prolongation} panel, as it sits between \textsc{Initial} and \textsc{Peak} (code wrapped around the conditional expression that would run if the expression evaluates to \code{True}).

\begin{figure}[htb]
    \centering
    \includegraphics[trim=0.2cm 0cm 0cm 0cm, clip=true, width=0.49\textwidth]{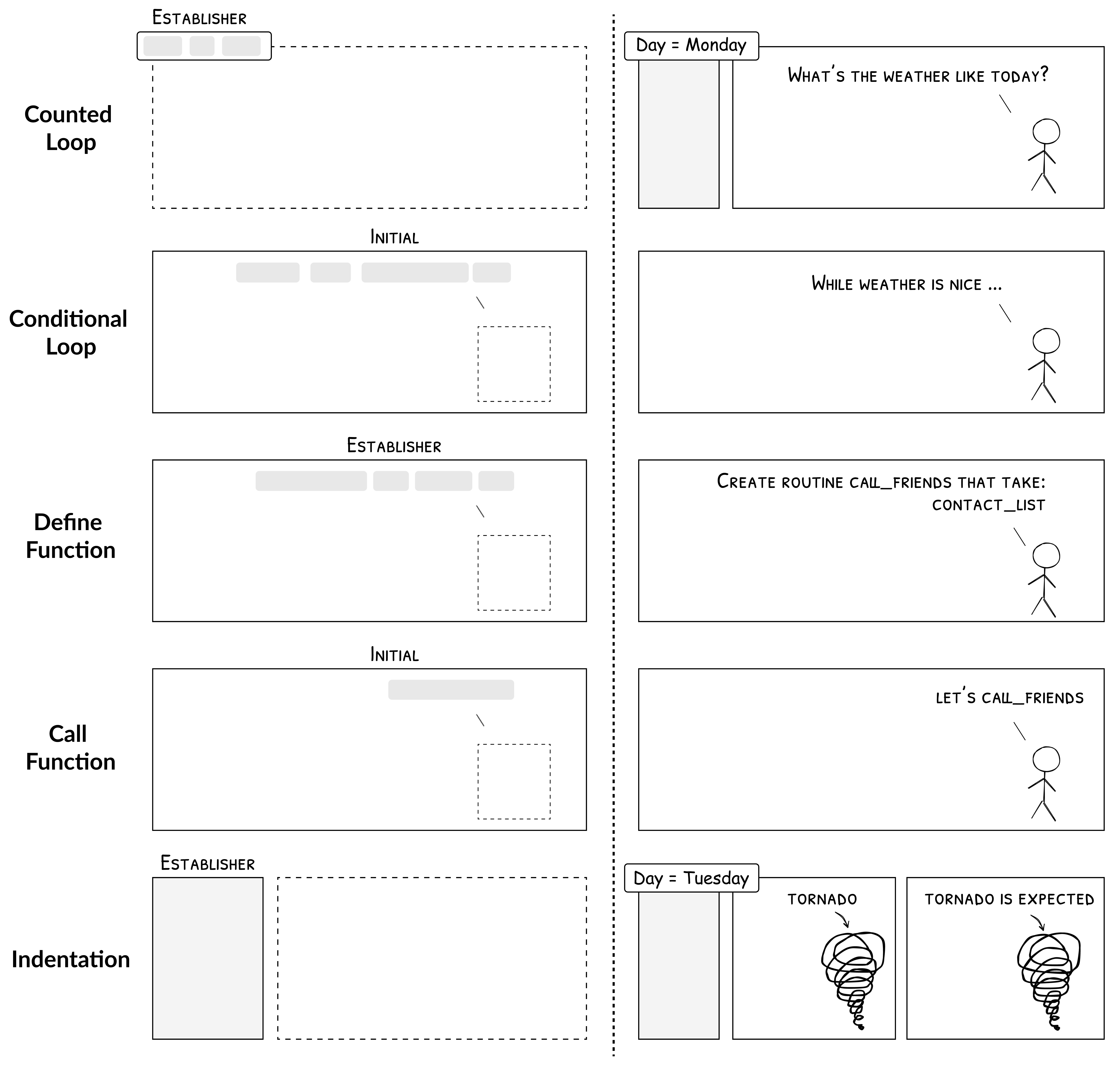}
    \caption{Other comic templates (left) and examples (right): the gray panels represent indentation; the dotted panels comic templates (e.g., variable assignment, counted loop); the gray boxes inside panels text (e.g., \textsf{Day = Monday}, \textsf{while weather is nice}); the dotted squares inside panels objects (e.g., stick figure, tornado, apple, banana).}
    \label{fig:visual_vocab}
\Description[Other comic templates for different code expressions.]{It shows other comic templates for different code expressions, which include counted loop, conditional loop, define function, call function, and indentation.}
\end{figure}

There can be more than one way to define these templates for coding strips depending on the goal(s), programming language and paradigm. In our case, the goals were: ensuring scalability, conveying semantics and executions, and offering design variations (e.g., if possible, avoid using only a single panel within a row); also, our visual vocabulary was created using Python (code expressions, syntax, and conventions). Understanding various nuances required to construct visual vocabularies for various programming languages is not within the scope of this research; we leave that as future work.

A key challenge in generating comics from code is maintaining a clear mapping between code, story, and comic (\textbf{D3}), so that learners can tell which line of code maps to which line of the story and which panel of the comic. We achieved this design goal (\textbf{D3}) by generating story template and comics that map 1-to-1 to lines of the code, as shown in Figs.\ref{fig:abstraction} and \ref{fig:structure_mapping}.
In addition to the 1-to-1 mapping, visual cues were also used to make the correspondence easily perceptible. For example, if there was an indentation in the code, this was carried over to the story template. For comics, an empty gray panel was used to indicate indentation, as shown in Fig.~\ref{fig:visual_vocab}. Note that, as shown in Fig.~\ref{fig:user-scenario}, when code has loop, the 1-to-1 mapping cannot be maintained as the comic needs to visualize the iterations.

\section{Evaluation} \label{evaluation}

To test whether CodeToon successfully supports the creative design process and generates quality comics from code, we conducted a two-part evaluation: a user study and a comic evaluation survey.

In the user study, we aimed to answer: \textbf{(RQ1)} Does CodeToon support the authoring of coding strip, in terms of story ideation and comic creation; \textbf{(RQ2)} Does CodeToon make the process of authoring coding strip more efficient; and \textbf{(RQ3)} What are the perceived utility and use cases of CodeToon for teaching and learning programming?
Note that \textbf{RQ3} does not focus on how well CodeToon supports the authoring of coding strip but on the perceived pedagogical value of CodeToon. While recent studies~\cite{suh2020coding, suh2021using} revealed learning benefits in participating in the process of creating coding strips and learning with coding strips, these studies were not done with CodeToon. This motivated us to explore \textbf{RQ3}, as it can reveal potential benefits and guide future work. Further, based on the comics created from the user study, in our comic evaluation study, we aimed to investigate: \textbf{(RQ4)} Does CodeToon help generate high-quality comics, and how consistent is the quality?

\subsection{Part 1: User Study (RQ1--RQ3)}

Our user study employed a between-subjects design with two conditions: Baseline (B) and CodeToon (C). Baseline was the same as CodeToon, but without the story ideation and auto comic generation features. Baseline users could generate story template from code. However, they did not have access to metaphor suggestions (Fig.~\ref{fig:structure_mapping}) and buttons (Fig.~\ref{fig:workflow_gc}) to instantly generate comics from the story template. Baseline users had to brainstorm story ideas on their own and manually create comics using the templates from the library (Fig.~\ref{fig:library}) and drawing tools (Fig.~\ref{fig:interface} (F\&G)).

\subsubsection{Participants}

We recruited all 24 participants (12 for each condition) from a local R1 university's study participant recruitment platform. Participants were required to have (1) basic programming knowledge, (2) a mouse, and (3) Chrome browser on their device.

Most Baseline users (age: M=23.3, SD=2.5; gender: 8F, 4M) had decent programming experience (6 Some Experience, 6 Much Experience), some experience with digital drawing tools (2 No Experience, 9 Some Experience, 1 Much Experience), and mostly positive perception towards the comics' usefulness as a tool for teaching and learning programming (1 Slightly, 1 Moderately, 6 Very, 4 Extremely Useful). Many were in the middle in terms of their confidence in drawing (4 Not Confident, 5 Somewhat confident, 3 Confident) and creating comics for teaching programming concepts (3 Not Confident, 8 Somewhat Confident, 1 Confident). They had varied experience with teaching programming (6 No Experience, 2 0-1 year, 4 1-3 years).

CodeToon users (age: M=27.3, SD=4.8; gender: 2F, 10M) also had decent programming experience (9 Some Experience, 3 Much Experience), some experience with digital drawing tools (3 No Experience, 7 Some Experience, 2 Much Experience), and similar perception towards the comics' usefulness (1 Slightly, 2 Moderately, 4 Very, 5 Extremely Useful). Many of them were not confident in drawing (7 Not Confident, 4 Somewhat Confident, 1 Confident) and creating comics for teaching programming concepts (8 Not Confident, 3 Somewhat Confident, 1 Confident). They had little to some experience teaching programming (3 No Experience, 5 0-1 year, 2 1-3 years, 1 3-5 years, 1 5+ years).

\subsubsection{Procedure} \label{study_procedure}

Baseline and CodeToon users followed the same study procedure. The study was conducted remotely using video conference software.
Participants first completed a pre-study survey. The survey included questions about their demographic information.
Then, they went through tutorial videos that explained the interface and how to use the tool, after which participants conducted practice tasks---replication tasks by following the videos. 
CodeToon users spent more time (at least 8 min) in the tutorial phase as they needed to learn and try story ideation and auto comic generation features.

Next, participants entered the task phase. The task was to create a comic about a programming concept of their choice. 
They were instructed that the end goal is to have code and corresponding comic that can be used together to teach students about the concept. Time limit was not imposed to study how users perform the task in a natural setting. Through pilot studies, we knew that participants generally had enough time to complete the task.

After the task phase, we administered three surveys: (1) post-study survey with creativity support index (CSI)~\cite{cherry2014quantifying}, (2) paired factor comparison (PFC)~\cite{cherry2014quantifying}, and (3) system usability survey (SUS)~\cite{bangor2008empirical}. 
We also conducted a short interview to ask participants to elaborate on their survey response to get a better understanding of what worked and what did not work. 
The study lasted between 1.5-2 hours and the participants received a \$30 Amazon gift card for their participation.

\subsection{Part 2: Comic Evaluation Study (RQ4)} \label{evaluation-comic_evaluation}

Our comic evaluation study followed the same procedure as in prior work~\cite{ngoon2021shown}, comparing comics resulting from two different conditions (Baseline and CodeToon) to understand whether CodeToon helped generate comics of better quality.

\subsubsection{Study Dataset}
After our user study, our comic dataset\footnote{Downloadable at \url{https://codetoon-research.github.io/download/}} consisted of 24 comics covering \textsc{variable} (1B, 0C), \textsc{condition} (7B, 7C), \textsc{loop} (2B, 2C), and \textsc{function} (2B, 3C).
We selected a subset instead of all comics for our survey for two reasons. First, it was not realistic to ask participants to rate all coding strips, as that can take approximately 2 hours (if 5 minutes per comic). We did not want to risk a survey receiving poor responses due to its length~\cite{galesic2009effects, kost2018impact}. Second, the quality of Baseline comics varied greatly. Since the amount of effort participants exerted ranged from very little to very high, 
we chose to select quality comics that participants put effort into (e.g., those who indicated they were satisfied with their work).
In this way, we could minimize variability and keep the survey at a reasonable length. 

Thus, we formed pair of comics (Baseline vs CodeToon) for the concepts by filtering for comics where participants answered in the post-study survey that they were `highly satisfied' with their comics (9 or 10 out of 10 on \textsf{`I was satisfied with what I got out of the tool'}), because participants' level of satisfaction varied (B: M=8.4, SD=1.6, Range=[6,10]; C: M=8.9, SD=1.3, Range=[6,10]). Then we grouped them into sets based on their concepts. This resulted in 2 sets for \textsc{condition}, 1 set for \textsc{loop}, and 1 set for \textsc{function}, in total 8 comics to rate for a 30 to 55-minute survey.

\subsubsection{Participants} 
We recruited 20 participants (age: M=24, SD=3.9) who can read and understand basic Python code through R1 university's study participant recruitment platform. Participants were mostly proficient in coding (1 Beginner, 1 Semi-Amateur, 4 Amateur, 8 Semi-pro, 6 Pro) and had moderate attitude towards comics' usefulness as a tool for teaching (2 Not At All, 3 Slightly, 8 Moderately, 4 Very, 1 Extremely Useful) and learning (3 Not At All, 1 Slightly, 9 Moderately, 4 Very, 2 Extremely Useful).

\subsubsection{Procedure} After signing up to participate, participants received the Qualtrics survey URL. After demographic questions, the survey presented eight comics in random order. Each page started with the programming concept the comic is based on, the comic, the code the comic is based on, and then a set of evaluation questions related to (1) how well the comic maps to code, (2) how well the comic illustrates code and concept, and (3) how useful they think the comic is for teaching and learning the concept it is based on. Participants received \$10 Amazon gift card for their participation.

\section{Results}

Baseline comics were generally inconsistent in their design language compared to CodeToon comics, as Baseline users manually designed their comics while CodeToon users incorporated auto generated comics. Thus, CodeToon comics were generally longer and more structured than Baseline comics. As for usability, both the Baseline and CodeToon users found the tool highly usable (SUS$_{B}$: 78.5, SD=20.5; SUS$_{C}$: 81.0, SD=9.6), giving usability scores well past the cutoff score (68) for production. For anonymity, we use B1{\ldots}B12 and C1{\ldots}C12 to refer to Baseline and CodeToon users, respectively.

\subsection{RQ1: Does CodeToon support the authoring of coding strips?}

\begin{figure}[htb]
    \centering
    \includegraphics[trim=0.2cm 0cm 0cm 0cm, clip=true, width=0.48\textwidth]{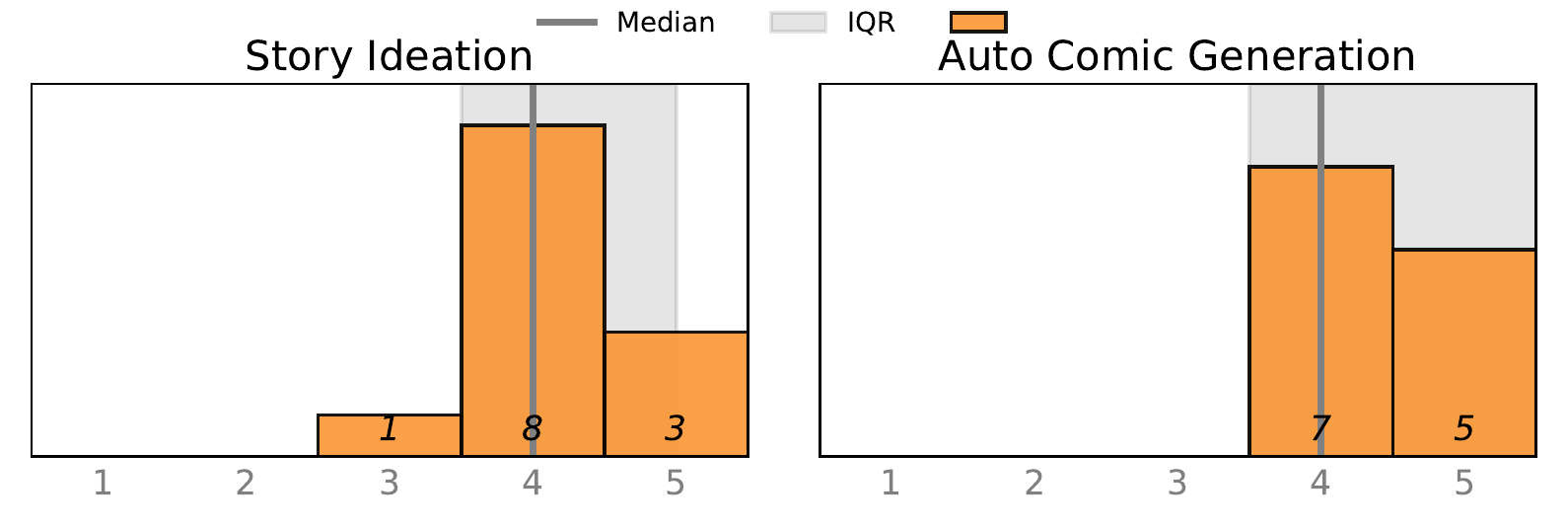}
    \caption{CodeToon (1: Not At All; 5: Extremely Useful)}
    \label{fig:post-study_features}
\Description[Two bar charts that show the perceived usefulness the two features, story ideation and auto comic generation.]{It shows two bar charts for the two features, story ideation and auto comic generation. The two charts show how 12 participants assessed the usefulness of the two features using the 5 Likert scale ratings, with 1 as not at all useful and 5 extremely useful. For story ideation, 8 participants gave 4, 3 participants gave 5, and 1 participant 3. For auto comic generation, 7 participants gave 4 and 5 participants 5.}
\end{figure}

We analyze a mix of responses from the survey, interview, and CSI results. We review the effectiveness of the proposed two features---story ideation and auto comic generation---as well as whether participants found the code-to-comic mapping accurate and the tool helped them be creative.
We begin with Fig.~\ref{fig:post-study_features}, which summarizes the participants' ratings on the usefulness of the two novel features.

\subsubsection{Feature 1. Story Ideation}
\label{codetoon_story_ideation-evaluation}

CodeToon users liked how story ideation helped ``spark creativity'' (C5) and ``think of creative ideas'' (C3). C6 noted that ``the options [in the dropdown]'' provided ``initial push of ideas'' to get started on figuring out what kinds of ``objects or things'' could be thought of. C8 was quite surprised that seeing the list of ideas could inspire him to create ``much more content beyond this object.'' C2 appreciated the fact that even though they had ``the extensive list of the objects that [they] could choose'' from, they could also write anything inside the input boxes (suggesting CodeToon satisfied one of our design goals, \textbf{D2}).

\subsubsection{Feature 2. Auto Comic Generation}

CodeToon users liked the auto comic generation feature because it saves ``a lot of time creating the layout'' (C12). C3 complimented CodeToon as a system that---even compared to other commercial comic drawing tools---is more useful, because it ``automatically generates the comics,'' decreasing the ``workload by a big factor.'' Several participants identified the auto comic generation as a novel feature, asking whether there has been work like this before.

\subsubsection{Impact of Two Features} 
\begin{figure}[htb]
    \centering
    \begin{subfigure}[t]{0.48\textwidth}
        \centering
        \includegraphics[trim=0.2cm 0cm 0cm 0cm, clip=true, width=\textwidth]{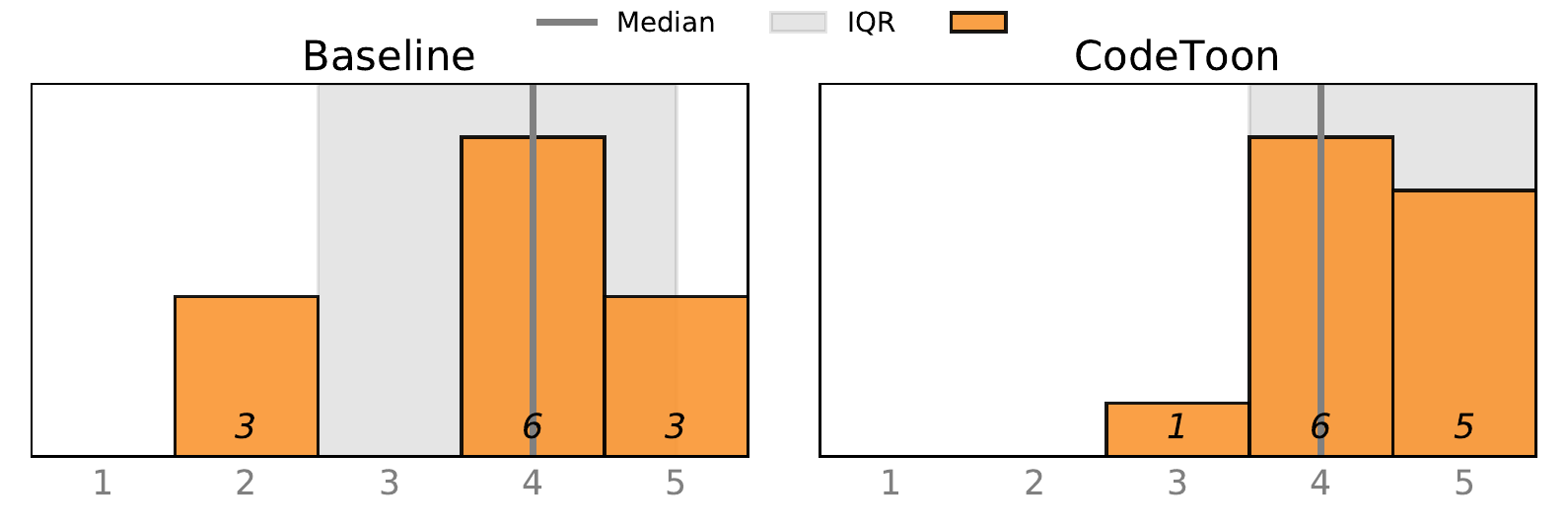}
        \caption{Usefulness (1: Not At All, 5: Extremely Useful)}
        \label{fig:post-study_comparison_useful}
    \end{subfigure}
    \begin{subfigure}[t]{0.48\textwidth}
        \centering
        \includegraphics[trim=0.2cm 0cm 0cm 0cm, clip=true, width=\textwidth]{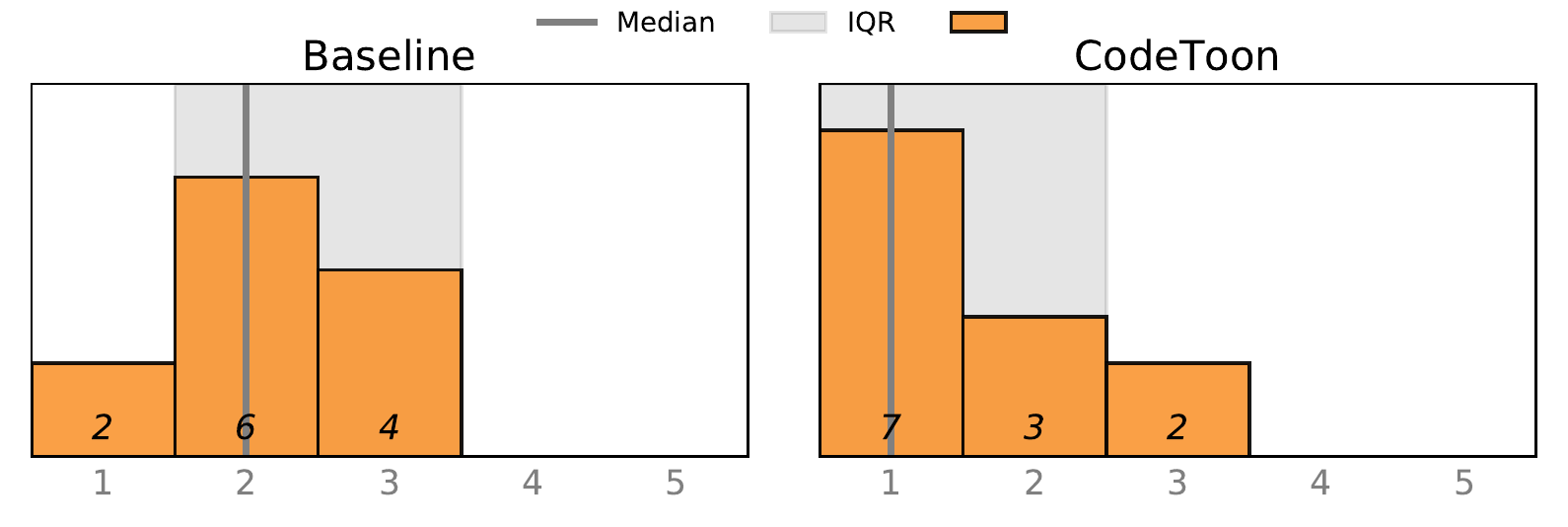}
        \caption{Difficulty (1: Not At All, 5: Extremely Difficult)}
        \label{fig:post-study_comparison_difficulty}
    \end{subfigure}
    \caption{Baseline vs CodeToon comparison on the (1) usefulness of the tool for the task and (2) the difficulty of the task with the tool. CodeToon users found the tool more useful for creating comics from code and the task less difficult.}
    \label{fig:post-study_comparison}
\Description[Two bar charts, one for the usefulness of the Baseline and CodeToon for the task of creating comics from code, and the other for the perceived difficulty of the task with Baseline and CodeToon.]{It shows two subfigures, (a) Usefulness and (b) Difficulty. (a) shows two bar charts for the usefulness of the Baseline and CodeToon for the task of creating comics from code. (b) shows two barcharts for the perceived difficulty of the task with Baseline and CodeToon. The bar charts are based on the 5 Likert scale ratings, with 1 as not at all useful and 5 extremely useful for (a) and not at all difficult and extremely difficult for (b). In the case of usefulness, CodeToon users rated slightly higher than Baseline users, with 9 of 12 participants answering 4 and 5 and 3 participants selecting 3. Nine Baseline users chose 4 and 5, with 3 participants selecting 2. In the case of perceived difficulty, CodeToon users found the task generally less difficult than Baseline users, with 7 participants answering not at all difficult (1) while 2 Baseline users answered not at all difficult and the rest chose 2 or 3.}
\end{figure}

Fig.~\ref{fig:post-study_comparison} provides further evidence that the two features facilitated the authoring of coding strips. Through the comparison, where the difference between CodeToon and Baseline is the presence of the two features, we see that they increased the perceived usefulness of the tool and decreased the perceived difficulty of the task.

\subsubsection{Code-to-Comic Mapping Accuracy.} 
\begin{figure}[hbt]
    \centering
    \includegraphics[trim=0.2cm 0cm 0cm 0cm, clip=true, width=0.48\textwidth]{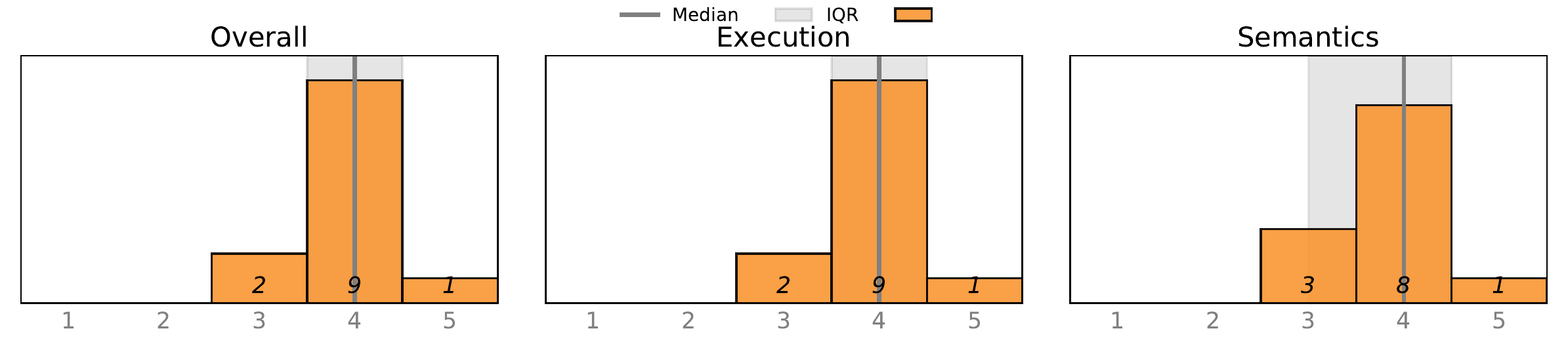}
    \caption{Mapping accuracy for auto generated comics. (1: Not At All, 5: Extremely Accurate)}
    \label{fig:post-study_accuracy}
\Description[Three bar charts on the mapping accuracy of the auto generated comics.]{It shows three bar charts on the mapping accuracy of the auto generated comics, with 1 as not at all accurate and 5 as extremely accurate. The three bar charts represent the overall accuracy, execution accuracy, and semantics accuracy. In all three bar charts, the accuracy was rated high, with participants answering 3,4, or 5. The overall accuracy and execution accuracy were the same, with 9 out of 12 participants answering 4. For the semantics accuracy, 8 participants answered 4.}
\end{figure}

To understand whether CodeToon users think that the generated comics illustrate code executions and semantics accurately, we asked them in the survey. As shown in Fig.~\ref{fig:post-study_accuracy}, they mostly found the generated comics' overall accuracy and their accuracy with the code executions and semantics very accurate. Many participants supported our design decision to map each line of the code to each row of the comic. C5 explained that she gave a high score to these accuracy questions because the generated comics mapped to the code line by line. C3 explained that this is why, compared to before the task, he feels comics are more useful. C3 said, ``I never thought [you] can teach the if conditions and conditional loops and everything to a student in the form of comics [in this manner].'' C5, who prior to using CodeToon thought about an alternative ``event-by-event approach,'' acknowledged that this design choice is better, saying: ``\ldots it is very helpful to have comics that go along with every single line just so that they can see what each line is actually doing. It is kind of like walking through the code step by step in a visual way\ldots I think it is better that it is line by line, because [it] makes it a lot easier for learners to make that connection.'' C8 made an interesting remark that accurate mapping between the comic and code can help address concerns about metaphors sometimes failing to communicate the ideas accurately and that if one were to use comics to teach programming, it has to be mapped to code like in CodeToon:

\begin{displayquote}
    ``I find that this tool maps the comic to the code very well\ldots more accurate than I have ever expected\ldots this balances the tradeoff between the barrier of learning the code and the inaccuracy of the metaphor. If teachers are to tell the student that the [comic] frames map to certain parts of the code, it has to be like this. Strictly mapped to the structure [of the code]\ldots this changed my thoughts on how helpful comic is.'' (C8)
\end{displayquote}

\subsubsection{Creativity Support Index}
As shown in Table~\ref{tab:csi}, CodeToon scored high and better than Baseline in all factors of the Creativity Support Index. There was a statistically significant difference in enjoyment, suggesting participants highly enjoyed using CodeToon. In fact, during the interview, many CodeToon users mentioned how fun the tool was. C1 said, ``This is a really interesting work. I really enjoyed playing with it.'' Moreover, all the CodeToon users except for one (11 More Confident, 1 Same as Before) who was already `Somewhat Confident' in creating comics about programming concepts before the task answered that they feel `More Confident' about creating comics about programming concepts after this task.

\begin{table}[hbt!]
\centering
\caption{Creativity Support Index Results. CodeToon performed better on every factor in creativity support. Statistical significance (\textit{p} < 0.05) is marked with $^{*}$.}
\label{tab:csi}
\begin{tabular}{l c c c}
  \toprule
  & Baseline & CodeToon & \textit{Sig.} \\
  \midrule
    Factor & Score (SD) & Score (SD) & \textit{p} \\ 
    \midrule 
    Enjoyment$^{*}$ & 15.9 (1.56) & \textbf{17.6} (2.0) & \textbf{0.01} \\
    Expressiveness & 16 (3.1) & \textbf{16.8} (1.5) & 0.21\\
    Exploration & 16 (3.1) & \textbf{16.4} (1.6) & 0.39 \\
    Immersion & 15.8 (4.1) & \textbf{16.8} (2.6) & 0.24 \\
    Results Worth Effort & 16.6 (2.4) & \textbf{17.8} (2.1) & 0.10\\
    \midrule
    Overall CSI Score & 80 (14.6) & \textbf{85.2} (7.5) & 0.14\\
     \bottomrule
  \end{tabular}
\Description[Creativity support index results.]{It shows a table for creativity support index results. In conclusion, CodeToon performed better than Baseline on every factor in creativity support. Among the five factors of the creativity support index, there is a statistically significant difference in one factor, Enjoyment (p=0.01).} 
\end{table}

\subsection{RQ2: Does CodeToon make the process of authoring coding strips more efficient?}

Our analysis showed that CodeToon users were able to save more than 6 minutes on average for the comic authoring (T-test: \textit{p}=0.08; Baseline: M=18:35, SD=11:17; CodeToon: M=11:45, SD=6:20) and the overall authoring time (Baseline: M=24:47, SD=13:43; CodeToon: M=18:27, SD=9:16). The average time spent on creating a story, on the other hand, was almost the same (T-test: \textit{p}=0.7; Baseline: M=6:11, SD=3:13; CodeToon: M=6:41, SD=3:38), which can be attributed to the relatively less time-consuming and less challenging nature of creating stories when compared to creating a comic. While CodeToon saved more than 6 minutes, the overall time difference was statistically insignificant (T-test: \textit{p}=0.19).

That CodeToon users spent less time is not surprising, given that CodeToon can instantly produce comics. After CodeToon users generated comics, they either submitted them as they are or spent some time simply adding a few details, e.g., additional panels before or after the generated comics to add additional context to the story. CodeToon users generally seemed satisfied with the layout of the generated comics. Although they were told they could edit them, none of them did. C3, who uses a PowerPoint in the workplace to explain code flow to coworkers, was impressed with how well CodeToon automatically generates a nice visualization to illustrate the code flow, saying ``[after CodeToon] automatically generates [comics,] the only thing [left to do] is the finishing touch.''

\subsection{RQ3: What are the perceived utility and use cases of CodeToon for teaching and learning programming?}

\subsubsection{Perceived Utility}
\begin{figure}[htb]
    \begin{subfigure}[t]{0.48\textwidth}
    \centering
        \includegraphics[trim=0cm 0cm 0cm 0cm, clip=true, width=\textwidth]{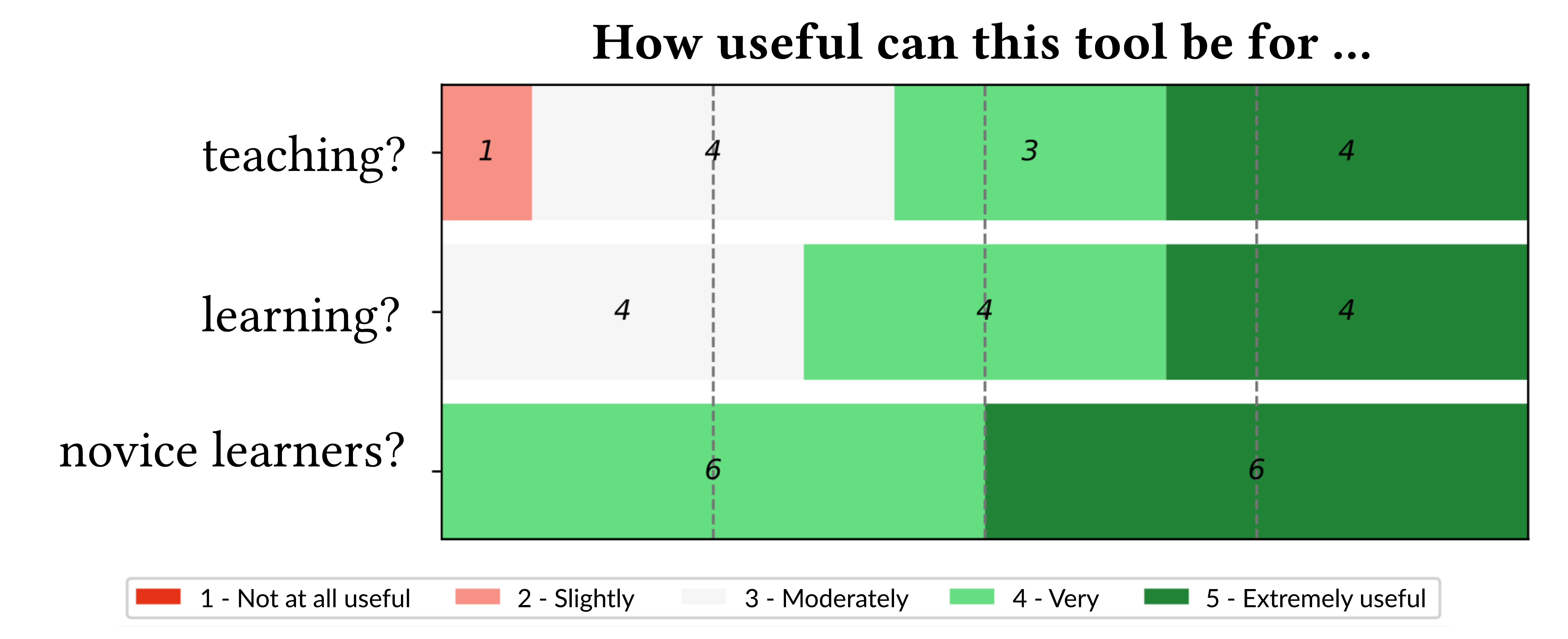}
        \caption{Baseline}
        \label{fig:baseline_utility}
    \end{subfigure}
    \begin{subfigure}[t]{0.48\textwidth}
    \centering
        \includegraphics[trim=0cm 0cm 0cm 0cm, clip=true, width=\textwidth]{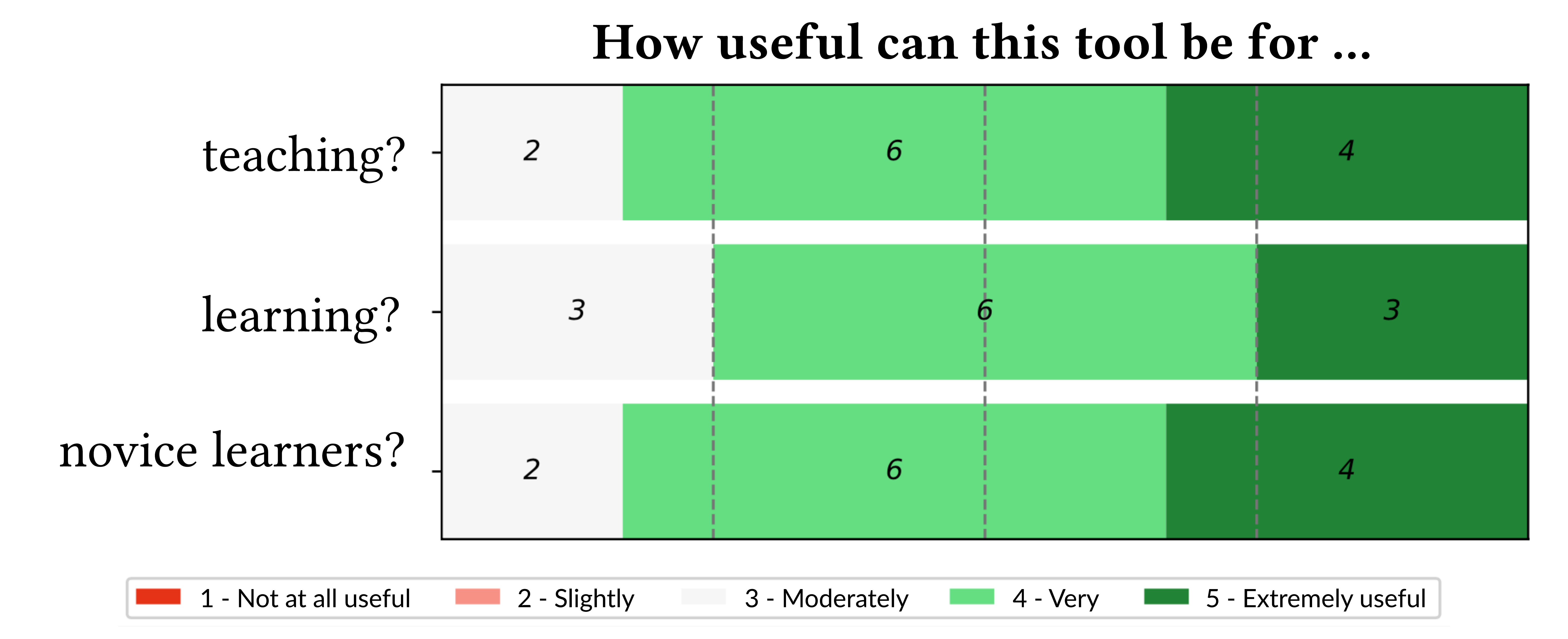}
        \caption{CodeToon}
        \label{fig:codetoon_utility}
    \end{subfigure}
    \caption{Perceived utility of Baseline and CodeToon for teaching and learning programming}
    \label{fig:perceived_utility}
\Description[Bar charts that show Baseline and CodeToon usersâ€™ answers to the question of how useful this tool can be for teaching, learning, and novice learners.]{It shows two subfigures, (a) Baseline and (b) CodeToon. Both (a) and (b) are bar charts that represent Baseline and CodeToon usersâ€™ answers to the question of how useful this tool can be for teaching, learning, and novice learners. The results are positive for both Baseline and CodeToon. Specifically, Baseline users mostly chose 3, 4, or 5 for the usefulness of the tool for teaching, learning, and novice learners. Only one Baseline participant chose 2 for the toolâ€™s usefulness for teaching. On the other hand, all CodeToon users chose 3, 4, or 5 for the toolâ€™s usefulness for teaching, learning, and novice learners.}
\end{figure}

As shown in Fig.~\ref{fig:perceived_utility}, most CodeToon users perceived CodeToon as a very and extremely useful tool for teaching, learning, and novice learners. Most of them (8/12) said they would like to use the tool for teaching; the others (4/12) who answered `Maybe' explained that this is because whether CodeToon is the right tool can depend on ``age,'' ``programming level,'' or ``learning styles.''\footnote{While we do not acknowledge the notion of ``learning styles,'' it has been added to avoid any loss of information or connotation.} Likewise, there were variations in the perceived utility of CodeToon. While several participants mentioned that the tool would be especially useful for ``younger students'' (C9), others suggested that it can be useful for older students (C8), such as undergraduate students (C5), and even for experienced programmers, e.g,. for ``debugging'' (C12). One thing everyone seemed to agree was that CodeToon provides a fun, creative way to learn programming.

\begin{figure*}[htb]
    \centering
    \includegraphics[trim=0cm 0cm 0cm 0cm, clip=true, width=\textwidth]{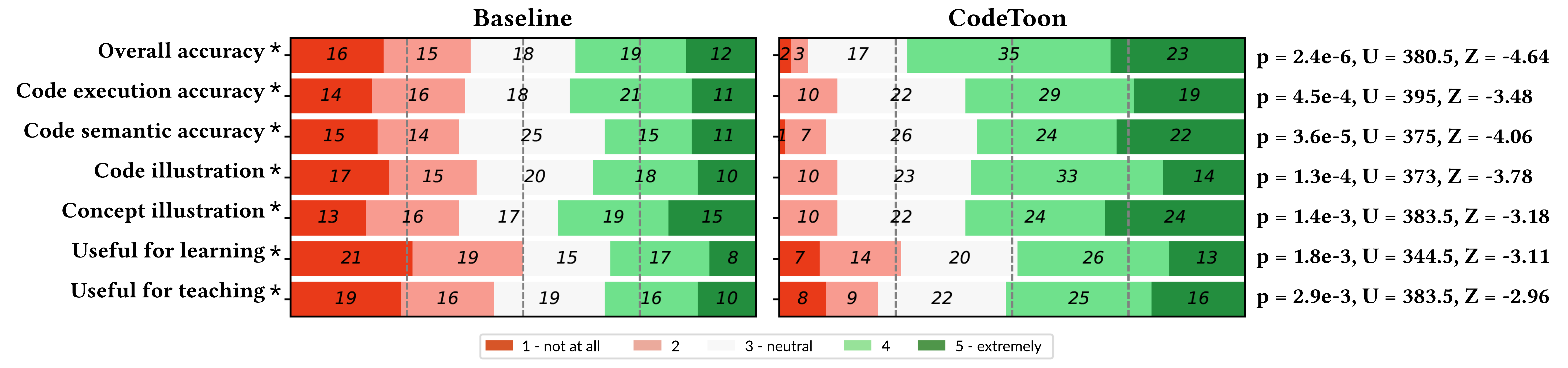}
    \label{fig:datacomicjs-interface}
    \caption{Comic evaluation results. Statistical significance (\textit{p} <0.05) is marked with *.}
    \label{fig:comic-evaluation-collective}
\Description[Two barcharts placed side by side for comparison of the usersâ€™ assessments of Baseline and CodeToon comics from the comic evaluation study.]{It shows two bar charts with Baseline and CodeToon placed side by side for comparison. The results show usersâ€™ assessments of Baseline and CodeToon comics from the comic evaluation study. The results show that there were significant difference between Baseline and CodeToon comics across all seven measures: overall accuracy, code execution accuracy, code semantic accuracy, code illustration, concept illustration, usefulness for learning, and usefulness for teaching.}
\end{figure*}

Participants provided several reasons for positively rating its utility. They thought that CodeToon can help students master ``important programming concepts,'' namely loop and function (C12). C12 said, ``especially in loop where [automatically generated comics] show \code{i=0}, \code{i=1}, \ldots comics give a visual explanation very well.'' C2 suggested that CodeToon would be useful for teaching data types to beginner students as it can show various examples (e.g., \textsc{integer}: \textsf{apple costs 10 dollars}; \textsc{boolean}: \textsf{apple tastes good}) to help develop intuitive understanding. C6 and C10 stated that the ``scaffolding'' in CodeToon would help students develop computational thinking, enabling them to see computational ideas in real-life situations and vice versa (i.e., moving between different abstraction levels, as shown in Fig.~\ref{fig:abstraction}). C6 noted that learning with CodeToon can show how ``computer science can be engaged beyond looking at codes on a screen\ldots [and that comic] is a very good lens to use.''

Most CodeToon participants (9 More Useful, 3 Same as Before) stated that the experience of using CodeToon made them realize that teaching/learning with comics (created from this tool) can be more useful than they had thought. 
Several participants explained that they had no idea that comics could be used in such a meaningful way. They explained that they thought comics were going to be used to merely narrate programming concepts. But seeing CodeToon generate comics that map to code line by line made them realize that comics can be ``very useful'' (C3).

\subsubsection{Use Cases.} \label{codetoon_usecases}

Participants provided several ways the tool can be used for teaching and learning programming. One group of related ideas was using the task as an exercise in ``classroom'' or ``tutoring settings'' as well as a group or individual assignment, after which they can ``present to class [their] story and understanding of the code.'' Another idea was holding a ``special programming class'' where students learn to read code after learning their corresponding comic expressions and then using it for exams or quiz where students---instead of explaining what a piece of code means---``can make comics to show what they mean.'' Participants also made several suggestions pertaining to how to use the tool. B6 suggested ``start[ing] with the comic first and then go[ing] to the story [and then] to the actual code,'' saying that ``this would help generalize the topic very easily.'' Several participants suggested using it to ``visualize'' basic programming concepts including ``data types'' as well as complicated concepts such as ``pointer'' (B12). Explaining how teachers utilize funny comic strips or memes in their teaching slides, participants also suggested that teachers can use the tool to quickly create and add visuals (comic) to their slides. C3 noted that the story ideation feature can also help teachers come up with appropriate examples and metaphors to explain the code and programming concept.

\subsection{RQ4: Does CodeToon help generate high-quality comics?}
\label{codetoon_rq4}

Fig.~\ref{fig:comic-evaluation-collective} shows how Baseline and CodeToon comics compare across the measures we investigated. As shown, there were statistically significant differences (Exact Wilcoxon-Mann-Whitney Test) between Baseline and CodeToon comics across all measures (accuracy, illustration, usefulness). In all measures, CodeToon comics were rated better than Baseline comics. That is, CodeToon comics were perceived as more accurate, illustrative, and useful for teaching and learning. Even in individual pair comparisons, CodeToon comics were also rated more positively (cf. Fig.~\ref{fig:individual-comparison} in Appendix).

Another important metric is whether CodeToon consistently produces high-quality comics. Along with the general observation that Baseline comics vary in quality while CodeToon provides comics of consistent quality (cf. Fig.~\ref{fig:individual-comparison}), responses to two agree/disagree statements from the CSI survey provide support. The two statements are: (1) \textsf{What I was able to produce was worth the effort I had to exert to produce it}; (2) \textsf{I was satisfied with what I got out of the tool}. In both statements (1 as Highly Disagree, 10 as Highly Agree), on average, CodeToon users ((1): M=8.83, SD=1.0; (2): M=8.92, SD=1.3) rated higher than Baseline users ((1): M=8.16, SD=1.5; (2): M=8.42, SD=1.6), suggesting that CodeToon users consistently produced comics of the quality they are satisfied with.

\section{Discussion}\label{discussion}

\subsection{Implications \& Opportunities}

\textbf{Code-Driven Storytelling.} The storytelling in CodeToon differs from prior work that leveraged storytelling in computer programming, as stories in CodeToon are structured around code. Whereas code has traditionally been used to define programmable aspects of stories such as animation and interaction (e.g., Scratch~\cite{resnick2009scratch}, Alice~\cite{kelleher2007storytelling}), in code-driven storytelling, code functions as a blueprint for stories: the three logic/control structures (sequential, selection, and iteration logic) in code become the structures of the story and comic. As it preserves the code's abstract structure across story and comic, learners can quickly recognize how they correspond to each other and make sense of code expressions, syntax, and conventions in the natural language and visual language of comics while engaging in the creative storytelling process. As general ideas in this approach will likely work with any programming language, this work opens up exciting opportunities to explore how we can leverage story ideation, auto comic generation, and structure mapping in a similar manner for other programming and computational languages (e.g., math).

\textbf{Storytelling with Text-based Programming.} CodeToon enables a way to learn computer programming through storytelling using text-based programming languages. Learning programming through storytelling has traditionally been done using block-based programming languages and environments~\cite{kelleher2007storytelling, resnick2009scratch}. Our computational pipeline for transforming text-based code to story and story to comic opens up a whole new direction to explore. For example, this approach could potentially enable us to teach young students who, due to their age, have first been introduced to coding via block-based programming languages. We could test whether they can learn with text-based programming languages using this approach or whether they can learn comic expressions and then use them to learn corresponding text-based programming expressions.
Since some students who learn coding first with block-based programming need to re-learn text-based programming when they grow older, this approach could potentially lessen this gap and provide an additional pathway to learning.

\textbf{Comics for Computational Languages.} 
The feasibility of our approach means that we can explore how other programming languages and paradigms can also leverage comics. For instance, object-oriented programming is an area where its concepts and code are often presented in terms of real-life equivalents~\cite{tanielu2019combining}. Comics could benefit students learning object-oriented programming. Other programming paradigms such as functional programming may require identifying abstractions (e.g., pointers) important to them for custom visual vocabulary, which would help expand the set of visual vocabularies and advance our understanding of mental models for programming.

\textbf{Design Implications.} Our work has interesting design implications for various domains. Our user study results suggest that the 1-to-1 mapping can be an effective, useful design when using comics as a complementary representation for code. This confirms the suggestion in the literature on multiple representational systems, which recommends making the mapping between multiple representations clear. For research areas---such as data and stats comics~\cite{bach2017emerging, wang2020data}---that also leverage comics in much the same way coding strip does, our insights concerning 1-to-1 mapping and the process of building comic expressions can help them explore a similar direction where they can also explore methods to auto generate comics from languages in their domain (e.g., R programming). Our work also leads to many interesting questions for comic authoring tools and coding tools. For instance, how can comic authoring tools utilize our idea of generating comics from language semantics and computational steps to enhance and diversify the authoring process? How can we design coding tools that offer ways to switch between code and other levels of abstraction (e.g., stories and comics)? How can we leverage such interaction to support various tasks programmers perform, e.g., reading and writing code, and debugging? What representations or abstractions can we support in these coding tools? How can we make their transitions seamless? How should we design these transitions and interactions so that we maximize the benefits of multiple representations?

\textbf{Visual Programming Environment for Artistic Activities.} 
Using coding strips to teach and learn computer programming is a new and promising approach. Recent work showed that it can enhance student learning and address some challenges in teaching programming~\cite{suh2020coding, suh2021using, suh2022phd}. But there is still much work to be done. To ease its adoption, we need a curriculum containing a set of learning activities and guidelines in an accessible form, such as a cheat sheet, so that teachers can quickly reference and apply it to their teaching and lessons. 
Moreover, understanding the nature and impact of this approach needs to be further investigated. For instance, what separates CodeToon from other visual programming environments like Scratch is that it can host artistic activities such as drawing. Recent efforts to combine art and programming---known as \textit{creative coding}---teach and use programming as the primary medium for creating visual artifacts. While CodeToon also allows users to do this to some extent with comic generation, its drawing canvas opens up opportunities for us to potentially explore a different direction---one that does not center around generating art with programming and allows students to learn computational ideas from artistic activities without having to first learn programming. This leads to several questions: What artistic activities can we develop? How can we incorporate them into teaching and learning programming? How can they improve teaching and learning in CS education?

\subsection{Limitations \& Future Work}

\textbf{Educational Benefits.} While prior work on coding strips demonstrated various learning benefits to using coding strips~\cite{suh2021using, suh2022phd}, learning with CodeToon is a different process that needs to be examined for several reasons. First, our findings on pedagogical usefulness are participants' \textit{perceived} usefulness. A rigorous study measuring its impact on learning would provide a more accurate assessment of its usefulness. Second, while our study participants included those with experience in teaching programming and learning as former students, their experience in programming and teaching varied. Finally, while CodeToon is a tool for teachers and students not yet proficient in programming, our user study participants were students with experience in programming. This was intentional, as we placed a higher priority on understanding possible usage scenarios and assessing the accuracy of the mapping between comics and code executions and semantics---which learners with little to no programming knowledge would be less qualified to do. Now that we tested CodeToon, we plan to do a study in the future with students who have little or no prior programming experience.

\textbf{Quality \& Limited Set.} For the drawings in the auto generated comics, we used the dataset from the Google's Quick, Draw! game~\cite{quickdraw_dataset}, which offers sketches of 345 categories. Since they are quick sketches people drew under pressure in a short time, the quality was mostly sub par. For better quality, we chose the AI correctly-guessed drawings. However, we saw several participants who were not satisfied switch to different objects. We suspect this may have given the impression for some that the tool is not for a professional audience but for a young audience. To address this quality issue and the challenge of having a limited set of categories, we plan to explore how we can leverage machine learning to convert existing pictures into comic-style sketches and then use them in the auto generated comics.

\textbf{Lack of Options.} Another limitation in CodeToon is a lack of support for different design languages and workflows. For instance, because we aimed to express semantics for every line, this led to auto generated comics illustrating code that is not executed. For example, for code inside the \code{if} block where the conditional expression evaluates to \code{False}, CodeToon still visualized this code even though it was not executed.
To indicate this, some participants suggested changing the opacity of these panels, whereas others suggested not showing them. While some were surprised they were shown (thus 4/5 instead of 5/5 for mapping accuracy on semantics), they recognized that this design would be helpful for teaching. Participants also shared alternative design ideas for auto generated comics---such as including code in the auto generated comic---and ideas for improving the workflow, such as real-time rendering. When a user adds a story to the story template, the comic is immediately rendered (or updated). Bi-directional mapping/rendering idea was also discussed---e.g., if the user edits comics, the story and code update accordingly in real-time. Suggestions like these show that users might desire different designs and workflows depending on the context. Therefore, we plan to implement a system preferences panel where users can customize, e.g., the design of the auto generated comics and rendering behaviors.

\textbf{Generative Models.} One exciting avenue to explore next is generative deep learning models.
Recently, large language models such as GPT-3 and Codex have shown impressive performance at language-related tasks, including generating code and translating natural language into code and vice versa. 
Generative model-based support can be an exciting addition to CodeToon. On user's request, generative conversational AI can generate or fill the gap in stories or code. Recent work showed that it can also generate stories from code or code examples for different programming concepts~\cite{suh2022leveraging}. Based on the analysis of the drawings on the canvas, it can also provide various creativity support, such as generating new drawings to add to the existing comic and providing design guidance in real-time.
For instance, if a user modifies the comic and it loses the relational structure carried over from code, it can notify the user and ask if that is the user's intention.
These generative model-based interactions would make the authoring process and learning experience more interactive, creative, and collaborative.

\section{Conclusion}\label{conclusion}

We introduce CodeToon, a comic authoring tool for facilitating the creation of coding strips by supporting story ideation and enabling auto generation of comics from code. To support this code-story-comic mapping, we used structure mapping theory and developed a visual vocabulary for coding strip. Then we tested whether CodeToon successfully supports the authoring of coding strips and helps generate high-quality comics through a two-part user study. The results of our user and comic evaluation studies show CodeToon not only supports the authoring of coding strips well but also reduces the comic authoring time by a significant amount while producing accurate, informative, and useful coding strips. In addition to contributing CodeToon and the two-part user study, our work lays the groundwork for code-driven storytelling by contributing computational pipeline and design guidelines for effective code-story-comic mapping.

\begin{acks}
This research was funded by the Learning Innovation and Technology Enhancement (LITE) Grant at the University of Waterloo. The authors also thank our study participants for their participation and reviewers for their feedback and suggestions.
\end{acks}

\bibliographystyle{ACM-Reference-Format}
\bibliography{references}

\appendix
\newpage
\onecolumn

\section{Appendix} \label{codetoon_implementation}

\begin{figure*}[htb]
    \centering
    \begin{subfigure}[t]{0.9\textwidth}
        \centering
        \includegraphics[trim=0cm 0cm 0cm 0cm, clip=true, width=\textwidth]{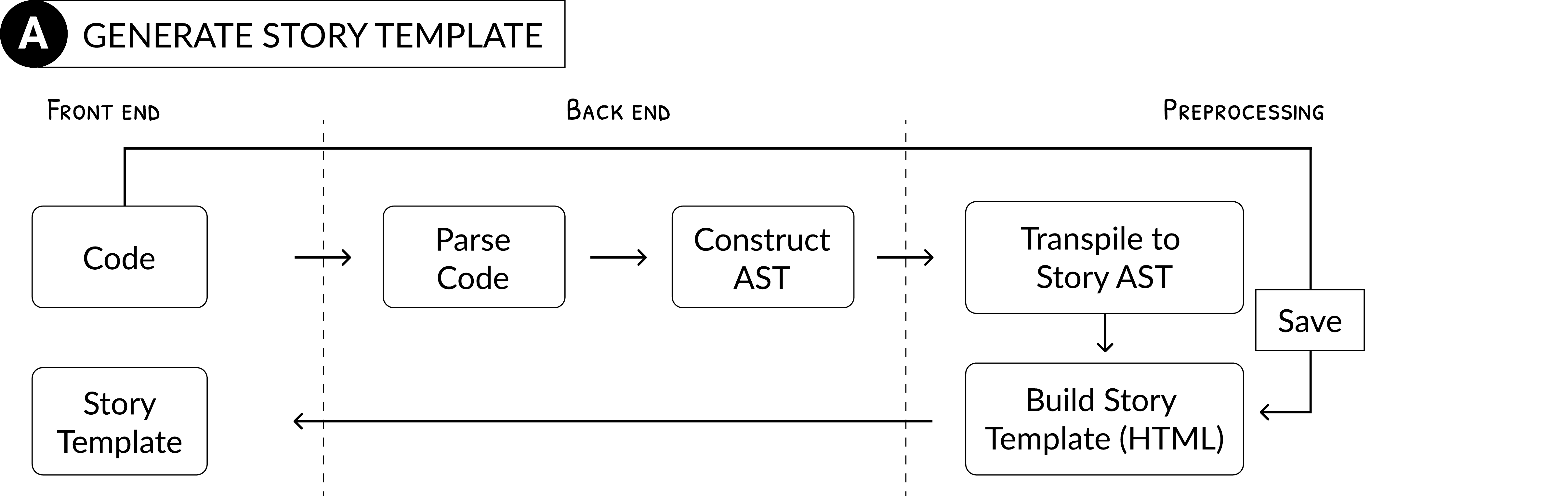}
        \caption{Pipeline for generating story template. When a user clicks the `generate story' button (Fig.~\ref{fig:interface}C), the code is sent to the back-end and parsed into an abstract syntax tree (AST) in JSON. To build a story template, the program traverses through the code AST (JSON) recursively. As it traverses, it checks node types (e.g., <Assign>) and extracts relevant information to build the story template, which consists mostly of \texttt{<input>} HTML tag. Once the story template has been generated, it is sent to the front end to be rendered in the story section (Fig.~\ref{fig:interface}D).}
        \label{fig:generate_story}
    \end{subfigure}
    \begin{subfigure}[t]{0.9\textwidth}
        \centering
        \includegraphics[trim=0cm 0cm 0cm 0cm, clip=true, width=\textwidth]{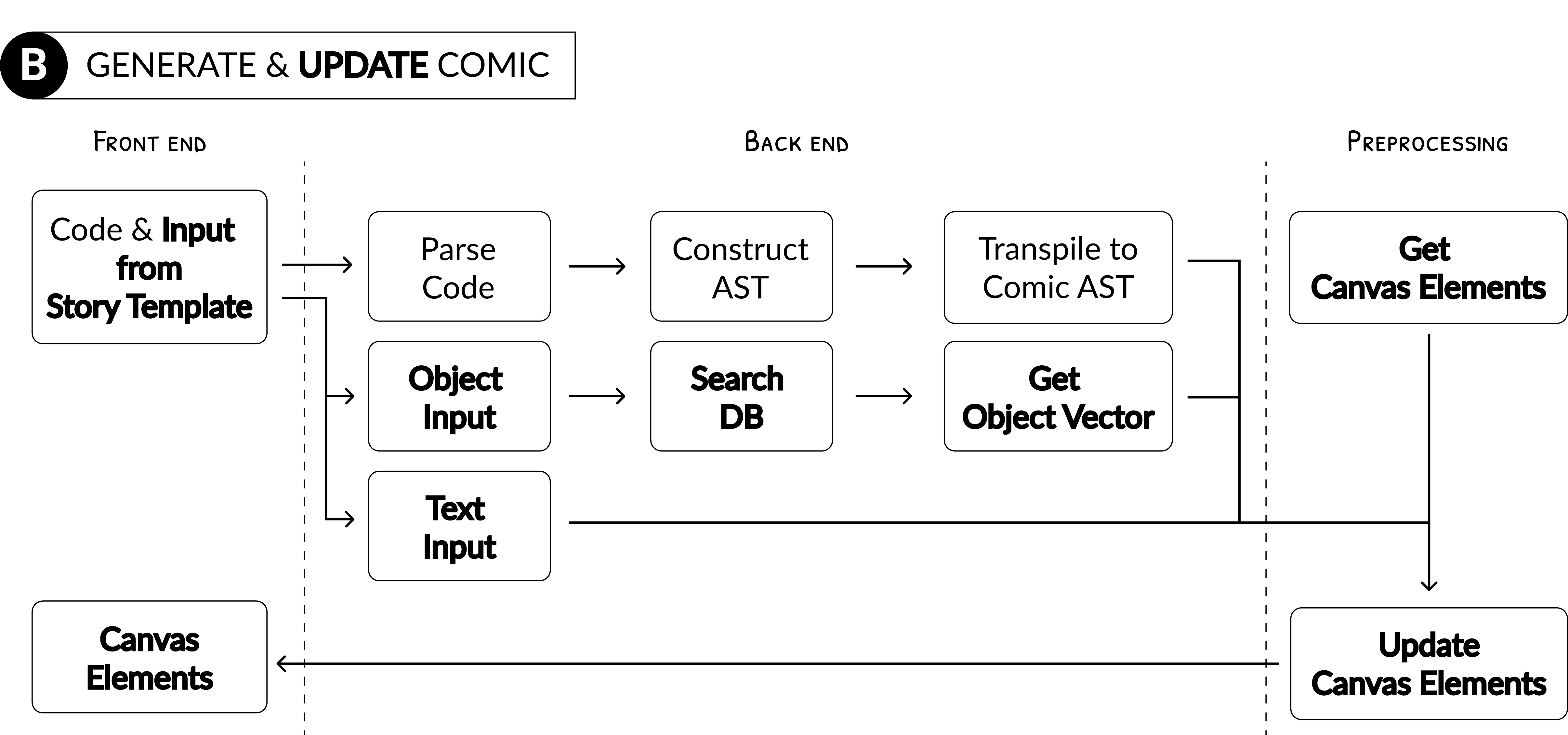}
        \caption{Pipeline for generating and updating comic from story. When a user clicks the `generate comic' button (Fig.~\ref{fig:workflow_gc}), CodeToon collects (1) code that was used to generate the story template and (2) values in the story template and makes an Ajax request. In the back-end, the program parses code and turns it into comic AST, which is essentially an instruction of how comic should be presented graphically (cf. Fig.~\ref{fig:autocomic_generation}). At the same time, the back-end checks whether any text made reference to the object in the object database. If there is a match, the back-end returns the vector information so they can be added to the comic. Once these information come together, they are sent to the front-end to render the comic in the canvas. As for updating the comic, when a user clicks the `update comic' button, it follows the same procedure (highlighted in bold), except that code is not sent to the back-end, because we are not generating a new comic. Note that `Get Canvas Elements' in the pipeline captures the elements currently present in the canvas. This is used to avoid (1) overwriting the entire canvas with new comic elements and (2) placing newly generated comic on top of existing elements.}
        \label{fig:add_update_comic}
    \end{subfigure}
    \caption{Implementation of computational pipeline for (a) generating story template and (b) generating and updating comic from story.}
    \label{fig:baseline-effort-examples}
\Description[Two flowcharts that explain how the two features in our system are implemented.]{This figure, found in the appendix of the paper, features two subfigures: (a) and (b). Both subfigures are flowcharts that explain the implementation of the computational pipeline for (a) generating story template and (b) generating and updating comics from story. The flowchart for the first subfigure (a) consists of 6 elements connected by action and flow links. It is as follows.
1. Code (in front-end).
a. Flows to 2.
b. Flows to 5.
2. Parse code (in back-end).
a. Flows to 3.
3. Construct code abstract syntax tree (in back-end).
a. Flows to 4.
4. Transpile to story abstract syntax tree (in back-end).
a. Flows to 5.
5. Build story template (in back-end).
a. Flows to 6.
6. Story Template (in front-end)

The flowchart for the second subfigure (b) consists of 10 elements. It is as follows.
1. Code & Input from Story Template (in front-end)
a. If the 'generate comicâ€™ button has been clicked, flows to 2.
b. If the update comic button has been clicked, flows to 5 and 8.
2. Parse Code (in back-end)
a. Flows to 3.
3. Construct code abstract syntax tree (in back-end)
a. Flows to 4.
4. Transpile to comic abstract syntax tree (in back-end)
a. Flows to 10.
5. Object Input (in back-end)
a. Flows to 6.
6. Search database (in back-end)
a. Flows to 7.
7. Get object vector (in back-end)
a. Flows to 10.
8. Text input (in back-end)
a. Flows to 10.
9. Get Canvas Elements (in back-end)
a. Flows to 10
10. Update Canvas Elements (in back-end)
a. Flows to 11.
11. Canvas Elements (in front-end)}
\end{figure*}

\begin{figure*}[h]
    \centering
    \includegraphics[trim=0cm 0cm 0cm 0cm, clip=true, width=\textwidth]{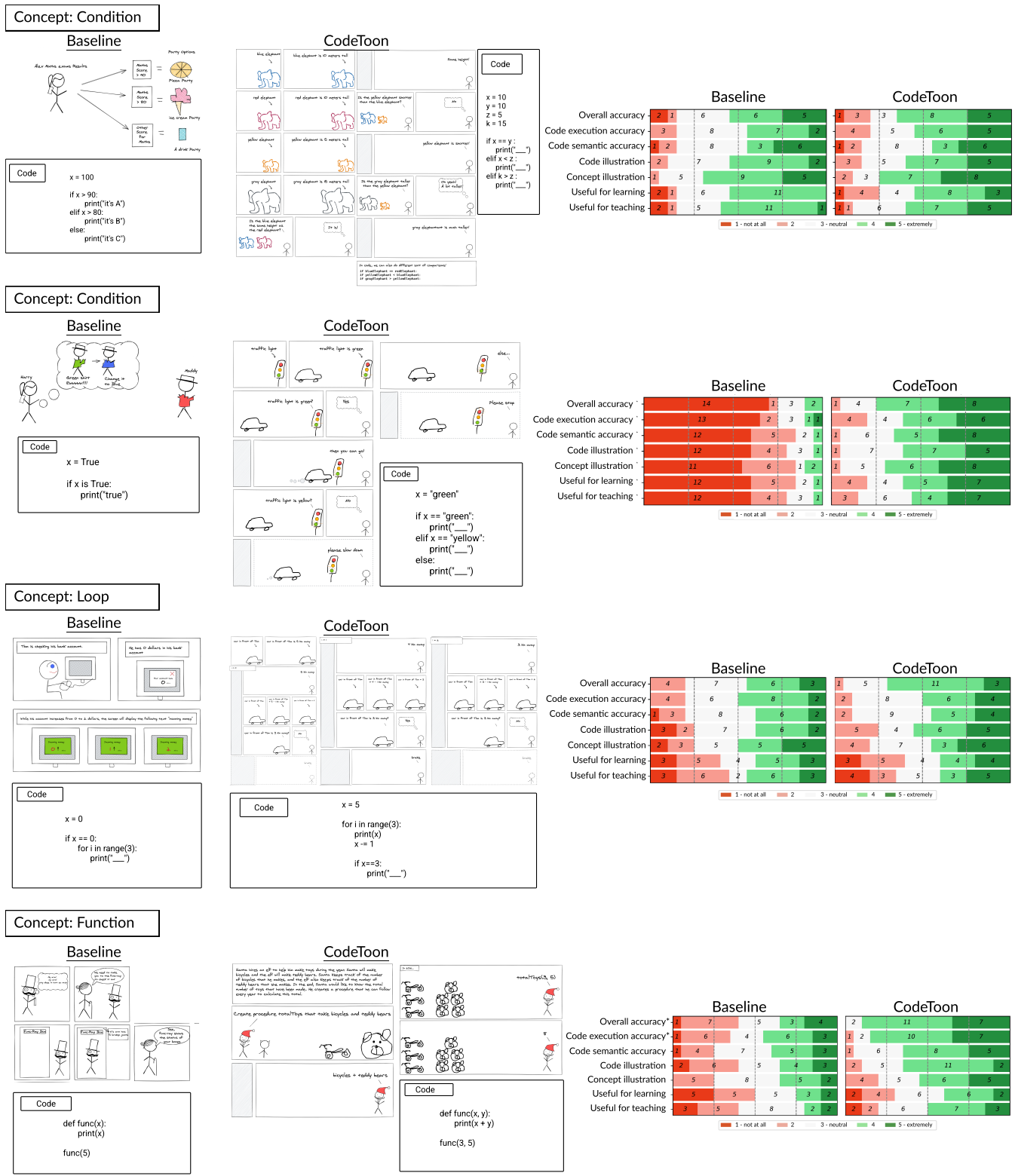}
    \caption{Comparisons of individual pairs according to their concepts. CodeToon comics were rated as being as good or better than Baseline comics in all cases across all measures.}
\Description[Comparisons of pairs of Baseline and CodeToon comics based on the same concepts.]{This figure, found in the appendix of the paper, shows comparisons of pairs of Baseline and CodeToon comics based on the same concepts. Baseline and CodeToon comics are shown on the left side of the figure, and the bar charts that reflect the assessment of these comics are shown on the right. Four pairs, which cover the concepts of condition, loop, and function, are shown.}
    \label{fig:individual-comparison}
\end{figure*}

\end{document}